\documentclass{article}

\usepackage[english]{babel}

\usepackage[letterpaper,top=2cm,bottom=2cm,left=3cm,right=3cm,marginparwidth=1.75cm]{geometry}

\usepackage{cite}
\usepackage{url}
\usepackage{amsmath}
\usepackage{graphicx}
\usepackage{pdfpages} 
\usepackage{float}
\usepackage{caption}
\usepackage{authblk}
\usepackage{parskip}
\usepackage[colorlinks=true, allcolors=blue]{hyperref}

\title{Bridging Technology and Humanities: Evaluating the Impact of Large Language Models on Social Sciences Research with DeepSeek-R1}
\author[1]{Peiran Gu}
\author[1]{Fuhao Duan}
\author[1]{Wenhao Li}
\author[1]{Bochen Xu}
\author[1]{Ying Cai}
\author[1]{Teng Yao}
\author[2]{Chenxun Zhuo}
\author[3]{Tianming Liu}
\author[1]{Bao Ge}
\affil[1]{School of Physics and Information Technology, Shaanxi Normal University, Xi’an, China}
\affil[2]{School of Foreign Languages, Northwest University, Xi'an, China}
\affil[3]{School of Computing, University of Georgia, GA, USA}
\begin{document}
\maketitle

\begin{abstract}
In recent years, the development of Large Language Models (LLMs) has made significant breakthroughs in the field of natural language processing and has gradually been applied to the field of humanities and social sciences research. LLMs have a wide range of application value in the field of humanities and social sciences because of its strong text understanding, generation and reasoning capabilities. In humanities and social sciences research, LLMs can analyze large-scale text data and make inferences. 

This article analyzes the large language model DeepSeek-R1 from seven aspects: low-resource language translation, educational question-answering, student writing improvement in higher education, logical reasoning, educational measurement and psychometrics, public health policy analysis, and art education. Then we compare the answers given by DeepSeek-R1 in the seven aspects with the answers given by o1-preview. DeepSeek-R1 performs well in the humanities and social sciences, answering most questions correctly and logically, and can give reasonable analysis processes and explanations. Compared with o1-preview, it can automatically generate reasoning processes and provide more detailed explanations, which is suitable for beginners or people who need to have a detailed understanding of this knowledge, while o1-preview is more suitable for quick reading.

Through analysis, it is found that LLM has broad application potential in the field of humanities and social sciences, and shows great advantages in improving text analysis efficiency, language communication and other fields. LLM's powerful language understanding and generation capabilities enable it to deeply explore complex problems in the field of humanities and social sciences, and provide innovative tools for academic research and practical applications. 
\end{abstract}

\hspace{2em} \textbf{Keywords:} LLMs, DeepSeek-R1, O1-preview, Humanities And Social Sciences

\newpage
\tableofcontents
\newpage
\section{Introduction}

In recent years, the application of Large Language Models (LLMs) in the humanities and social sciences has developed rapidly~\cite{zhong2024opportunities,huang2024social}. This paper aims to evaluate the performance of DeepSeek-R1~\cite{guo2025deepseek} and o1-preview~\cite{zhong2024evaluation} in some humanities and social sciences fields, namely low-resource language translation, educational question answering, student writing improvement in higher education, logical reasoning, educational measurement and psychometrics, public health policy analysis, and art education.

The o1-preview model released by OpenAI is known for its reasoning ability and ability to solve complex problems, especially in mathematics, programming, and science. The release of this model is seen as an important step towards AI with human cognitive capabilities.

The advantage of DeepSeek-R1~\cite{mercer2025brief} is more efficient reasoning and cost control, that is, through lightweight architecture and model structure optimization, it significantly reduces computing resource consumption while maintaining high performance, faster response speed, suitable for concurrent scenarios, and lower unit computing cost, providing users and enterprises with more cost-effective services, optimizing the ability to understand long texts, and being able to track user questions and answers more coherently.

\subsection{DeepSeek-R1 Brief Introduction}

DeepSeek-R1 is a large language model developed by China's DeepSeek company. Its design goal is to provide efficient, reliable and secure intelligent interactive services, including knowledge question and answer, task planning and other scenarios.

DeepSeek-R1 is based on the Transformer~\cite{vaswani2017attention} architecture and uses a self-attention mechanism, which enables the model to capture global information more efficiently when processing sequence data. The model architecture uses a variety of optimization methods such as hybrid expert architecture, multi-head potential attention mechanism, auxiliary loss load balancing, multi-Token prediction, FP8 mixed precision training, etc~\cite{guo2025deepseek}. In terms of model training methods, knowledge distillation, reinforcement learning, multi-stage training and cold start data are used to reduce computing requirements and costs while ensuring the accuracy of the model.

\subsection{O1-preview brief introduction}

O1-preview is a large language model released by OpenAI in September 2024, which aims to improve the performance of artificial intelligence in complex reasoning tasks. Compared with the previous GPT series models~\cite{floridi2020gpt,kublik2022gpt,achiam2023gpt}, o1-preview has complex reasoning capabilities, such as: o1 will generate a series of internal thought chains before answering questions to simulate the human thinking process, so as to understand the problem more deeply and provide accurate answers. Through reinforcement learning, o1 optimizes its thinking process. o1 performs well in benchmarks in multidisciplinary fields such as chemistry and physics, reflecting major progress in AI technology. Although o1 demonstrates strong reasoning capabilities, its longer reasoning time can lead to slower response speeds in some cases.

\subsection{Research Objectives}

The research objectives of LLM in the field of humanities and social sciences mainly include improving language and text analysis capabilities, such as natural language understanding, text generation, etc., helping to analyze literature and language translation, etc. In the field of education, LLM can be used to assist students in answering questions, provide analytical steps for answering questions, etc., optimize learning experience through personalized education and cognitive research, and promote thinking research. At the same time, LLM can also assist in policy research and analyze the impact of legal policies on society.

This paper will compare and analyze DeepSeek-R1 and o1-preview in seven humanities and social sciences related fields: low-resource language translation, student writing improvement in higher education,educational Q\&A, logical reasoning, educational measurement and psychometrics, public health policy analysis, and art education. In each field, multiple questions and answers are used to test and evaluate the accuracy and effectiveness of the two models in these fields, including the model's reasoning ability, language expression ability, question-and-answer logic, etc.

\newpage

\section{Research content and datasets}

This section mainly describes the main contents and data sets of the seven research aspects in the selected humanities and social sciences fields, and briefly describes the experimental content and results.

\subsection{Low-Resource Language Translation}

Current large language models (LLMs) have mediocre results for low-resource language translation tasks and perform poorly in some scarce resource languages~\cite{zhong2024opportunities,merx2024low}, such as data scarcity, complex language structure, and high requirements for model generalization capabilities. Large language models usually have high requirements for training data. Low-resource language translation tasks usually lack large-scale, high-quality bilingual corpora, which makes it difficult for models to effectively learn and capture features. In addition, some low-resource languages have complex grammar, with many changes in word affixes and grammatical rules that differ greatly from high-resource languages, making data sparsity more serious and increasing the difficulty of model learning. In terms of language alignment, most low-resource languages have less aligned corpora, which affects the model's alignment learning ability.

This section uses the data from the main part of the Cherokee corpus (Cherokee-English
Dictionary)~\cite{nuttlecherokee} to evaluate DeepSeek-R1's ability to handle low-resource language translation and compare it with o1-preview's ability to translate low-resource languages.Cherokee-English Dictionary (CED) is a language resource project aimed at preserving and promoting the Cherokee language. It builds a Cherokee-English dictionary to help researchers and Cherokee members better understand and use Cherokee, which is a critically endangered language currently used mainly by a small number of Cherokee elders and meets the criteria for a low-resource language.

Experimental results show that both DeepSeek-R1 and o1-preview have the ability to recognize words and basic grammar in low-resource language translation, but they encounter greater challenges in translation accuracy when dealing with more complex language scenarios. The two models have their own strengths in translating low-resource languages, and the difference between them is not large. The difference in translation depends on the given context and the required translation accuracy requirements.

\subsection{Student Writing Improvement in Higher Education}

This section focuses on assessing Deepseek-R1's potential to improve academic writing competencies in university settings. Higher education writing demands mastery of disciplinary conventions, logical organization, and context-aware communication - areas where Deepseek-R1's sophisticated linguistic features may provide targeted support. Unlike conventional LLMs that primarily address grammatical correctness, our evaluation framework examines five critical dimensions: textual precision, argument flow optimization, research blueprint development, source integration, and adaptive writing styles. The assessment utilizes authentic student compositions from the Corpus \& Repository of Writing (CROW)\cite{StaplesDilger2018}, comprising multi-stage drafts across diverse academic disciplines and proficiency levels from writing-intensive courses at three U.S. institutions. This stratified dataset enables nuanced examination of the tool's effectiveness in addressing varied pedagogical needs throughout the writing process.

The results show that the content generated by DeepSeek-R1 shows obvious advantages in discourse depth, terminology precision and emotional expression, while the o1-preview output tends to be mechanical theoretical analysis, which is characterized by a limited reference basis and reduced contextual adaptability.

\subsection{Educational Q\&A}

Educational Q\&A is a combination of deep learning with natural language processing and information retrieval. It aims to answer various subject-related questions raised by users. Such systems can not only provide answers, but also generate detailed explanations and examples and recommend learning resources, which helps improve user learning efficiency. Large language models are widely used in the field of educational question-answering, mainly covering knowledge question-answering, knowledge retrieval and recommendation in various subjects.

This section focuses on large language models in the field of educational question answering, mainly studying efficient reasoning, language logic, etc. DeepSeek-R1 optimizes the question-answering model for educational scenarios. It is trained through high-quality educational data such as textbooks, test questions and other texts to improve the depth and accuracy of subject knowledge coverage. For example, the complex reasoning ability in fields such as geography and physics is significantly improved. The knowledge completeness of the model is improved through the fusion of retrieval enhancement generation and knowledge graph, so that it can accurately quote academic resources such as textbooks to answer. DeepSeek-R1's innovations in technical architecture, such as the application of hybrid expert architecture, dynamically activate different expert modules to ensure response speed while improving the processing accuracy of complex problems. In response to the real-time needs in the field of educational question answering, DeepSeek-R1 uses pruning, quantization and other technologies to reduce computing costs without affecting the accuracy of answering questions.

The research data in this paper uses the SciQ dataset~\cite{Welbl2017}, which contains more than 13,000 high-quality choice questions in various subjects such as geography and physics. This paper uses several questions selected from this dataset to evaluate the ability of the DeepSeek-R1 model in understanding and reasoning knowledge in various subjects and compares it with the o1-preview model.

The results show that DeepSeek-R1 performs well in the field of educational question answering, demonstrating the powerful capabilities of large language models in the field of educational question answering. The model has a deep understanding of multidisciplinary concepts and can avoid misleading interference factors to choose the correct answer. Compared with the answers of the o1-preview model, DeepSeek-R1 often gives more detailed answers, which is suitable for detailed learning.

\subsection{Logical Reasoning}

In the era of artificial intelligence, logical reasoning stands as a cornerstone capability distinguishing human cognitive intelligence from superficial pattern matching. While large language models have made strides in natural language understanding, their limitations in complex logical reasoning—such as multi-step deductions, modal logic processing, and cross-domain transferability—remain a critical bottleneck.For LLMs, logical reasoning ability directly affects their performance in several key tasks such as text understanding and generation and problem solving. The ability of large language models to judge the causal relationship between sentences through reasoning is the key to generating more coherent and logical text. In the fields of legal analysis and language translation, reasoning ability determines whether the model can give reasonable answers, rather than just searching the existing knowledge base.

This section uses choice questions from the \href{https://github.com/lgw863/LogiQA-dataset}{LogiQA dataset}~\cite{LgwLogiQA} to test DeepSeek-R1 and compares the test results with o1-preview. This dataset is designed to evaluate the performance of natural language processing systems in logical reasoning tasks. It contains more than 8,000 examples, covering various types of deductive reasoning, presented in the form of choice questions. Currently, most large language models perform far below human level on this dataset, so this dataset can be used as a benchmark for studying logical large language models in the context of deep learning natural language processing. In addition, the dataset also contains diverse contextual information and background knowledge to simulate real-world human logical reasoning scenarios.

The experimental results show that DeepSeek-R1 has divergent ideas in logical reasoning, but its accuracy is slightly worse than that of the o1-preview model.

\subsection{Educational Measurement and Psychometrics}

The integration of large language models (LLMs) into educational measurement and psychometrics is emerging as a critical advancement in assessment design, item response analysis, and adaptive testing. This section conducts a comparative evaluation of two prominent LLMs, o1-preview and DeepSeek-R1, in their ability to analyze psychometric models and their applications in large-scale educational assessments. By systematically comparing their performance, we aim to elucidate their respective strengths and limitations in addressing complex measurement challenges.  

This evaluation is structured around three fundamental dimensions of psychometrics: the validity and reliability of standardized tests, the effectiveness of adaptive testing methodologies, and the impact of measurement error on educational outcomes. These issues are central to both theoretical research and practical implementation in educational assessment. To ensure a rigorous evaluation, we selected three research questions from \textit{Advances in Educational and Psychological Measurement: Theories and Applications}\cite{hambleton2013advances} as the basis for our analysis. The models' responses were assessed using a structured question-and-answer framework, allowing for a systematic examination of their ability to handle real-world psychometric challenges.  

To compare the performance of DeepSeek-R1 and o1-preview, we designed three prompts that reflected authentic educational measurement scenarios, requiring precise, evidence-based responses. The generated outputs were then evaluated against established theoretical perspectives and expert assessments to measure their accuracy, analytical depth, and consistency.  

Despite the limited scope of this evaluation, the findings suggest that both o1-preview and DeepSeek-R1 exhibit substantial potential in psychometric analysis. Each model demonstrated distinct strengths: o1-preview excelled in statistical modeling and predictive analytics, while DeepSeek-R1 proved more adept at synthesizing theoretical constructs and proposing innovative test designs. These results offer valuable insights into the future application of LLMs in educational measurement and psychometrics, highlighting their complementary capabilities in advancing the field.  

\subsection{Public Health Policy Analysis}

The use of large language models (LLMs) in public health is emerging as an important tool to enhance health policy making, healthcare delivery management, and disease surveillance. This section will take an in-depth look at the performance of two leading LLMs, o1-preview and DeepSeek-R1, in analyzing the Affordable Care Act (ACA)~\cite{USDHHS2024}. By comparing the performance of these two models, we hope to reveal their unique strengths and potential limitations when dealing with complex health policy issues.

This section focuses on three core areas: insurance coverage expansion, the impact of reforms on chronically ill populations, and differences in Medicaid programs across populations. These issues not only represent key achievements in the decade of ACA implementation, but also reflect the most pressing needs in current health policy. We took three specific questions from the article named The Affordable Care Act at 10 Years: Evaluating the Evidence and Navigating an Uncertain Future as the basis for our evaluation and tested them using a question-and-answer format. This approach allowed us to meticulously assess how well each model performs in the face of real health policy challenges .

To evaluate the performance of DeepSeek-R1 and o1-preview, we designed three prompts based on real-world data that required the models to provide concise, accurate answers. We compared the generated answers with the views of domain experts to assess the accuracy, depth, and consistency of the models.

Despite the limited size of the dataset used for this evaluation, the results show that both DeepSeek-R1 and o1-preview have significant potential for health policy analysis. Each demonstrated different strengths, for example, o1-preview was better at analyzing data and predicting trends, while DeepSeek-R1 was better at synthesizing information from multiple sources and proposing innovative solutions. This provides a valuable reference for future applications.

\subsection{Art Education}

This section evaluates the performance of the DeepSeek-R1 and o1-preview models in the field of art education, focusing on their capabilities in understanding theoretical curriculum concepts and planning art-related instructional activities. By comparing their responses to those of human experts under identical problem scenarios, we examined their potential to assist human educators in this domain.

We specifically designed two core tasks: 1) interpret theoretical frameworks in art education (for example, explaining William Pinar's concept of currere~\cite{Pinar2019}), and 2) plan practical art curriculum activities (for example, design a cardboard assemblage art activity~\cite{Penfold2020}). These tasks were grounded in real-world educational contexts to test the models' conceptual comprehension and practical curriculum design abilities.

Overall, the research highlights the comparative strengths and limitations of DeepSeek and o1-preview, exploring how cutting-edge AI models differ from human educators in art education. It further investigates the potential of these models to serve as tools for art educators, aiding in theoretical instruction and activity planning while identifying areas where human expertise remains irreplaceable.

\newpage

\section{Related Work}

This section mainly introduces the relevant architecture and technology of DeepSeek-R1~\cite{guo2025deepseek}.DeepSeek-R1 can achieve low cost with high accuracy, which lies in the innovation of model architecture design, training strategy, etc. In terms of model architecture, DeepSeek-R1 and o1-preview are both based on transformer architecture~\cite{vaswani2017attention}. DeepSeek-R1 adopts MoE module~\cite{cai2024survey} and built-in chain reasoning method based on transformer architecture. In terms of training, pure reinforcement learning and multi-stage training pipeline are used. The optimization methods in training include knowledge distillation and mixed precision reasoning. Through these methods, DeepSeek-R1 achieves a balance between cost and efficiency.

\subsection{Reinforcement Learning}

DeepSeek-R1 attempts to use pure reinforcement learning to improve the reasoning ability of language models and study the potential of LLM to develop reasoning ability without any supervised data~\cite{guo2025deepseek}. The core of reinforcement learning is the Markov decision process~\cite{kaelbling1996reinforcement}. The model continuously tries and learns by interacting with the environment to maximize long-term rewards. It usually evaluates the quality of each state through a value function. The key steps of reinforcement learning is to continuously optimize strategies based on trial and error, which is suitable for long-term decision-making and complex environments~\cite{wiering2012reinforcement,li2017deep}. DeepSeek-R1 uses GRPO~\cite{mroueh2025reinforcement} as an RL framework to improve the performance of the model in reasoning~\cite{guo2025deepseek}.

In reinforcement learning, cold start and multi-stage training pipeline are two important concepts, which respectively solve the problem of the model's lack of experience in the initial stage and how to train efficiently.
\subsubsection{Cold start Problem}

The cold start problem~\cite{ding2017cold,ji2021reinforcement} refers to the fact that the AI model has no experience at first and does not know how to choose the best action, which leads to problems such as high random exploration cost, slow convergence, and sparse rewards. In complex environments, the model may not receive positive rewards for a long time, resulting in learning difficulties. During the training process, DeepSeek-R1~\cite{guo2025deepseek} collected thousands of cold start data to fine-tune the DeepSeek-V3-Base model~\cite{liu2024deepseek}, then it performed RL for reasoning.

\subsubsection{Multi-Stage Training Pipeline}

The multi-stage training pipeline in reinforcement learning is a method to gradually improve the performance of the agent through a phased training strategy~\cite{guo2025deepseek,schnabel2025multi}. This method decomposes complex tasks into multiple subtasks or stages, each stage focusing on a specific subtask, thereby gradually guiding the agent to learn from simple to complex and from local to global, thereby improving training efficiency and final performance. The training process of DeepSeek-R1 is mainly divided into four steps, namely cold start, reasoning-oriented reinforcement learning, rejection sampling and supervised fine-tuning, reinforcement learning in all scenarios, and gradually building a general and reliable model capability. Its core advantage lies in the phased training objectives to improve model performance, efficiency and generalization ability.

\subsection{DeepSeek-R1 Optimization Method}

\subsubsection{Knowledge Distillation}

Hinton proposed the idea of distillation in 2015~\cite{hinton2015distilling}, with the main goal of improving the performance of machine learning algorithms without increasing computational complexity. Knowledge distillation is to guide the student model through the output generated by the teacher model, so that it imitates the teacher's generation style and content quality to maintain the fluency and relevance of the generated content~\cite{mcdonald2024reducing,xu2024survey,latif2023knowledge}. DeepSeek-R1 focuses on retaining the ability to understand the semantics, grammar and cultural context of some specific problems during the distillation process, and conducts targeted training for specific tasks~\cite{guo2025deepseek}.

\subsubsection{Mixed Precision Inference}

The advantage of mixed-precision reasoning technology is that it balances computational efficiency and model accuracy~\cite{chen2024progressive,zheng2024mixllm}. DeepSeek-R1's mixed-precision reasoning technology integrates different numerical precisions~\cite{guo2025deepseek}, significantly improving reasoning efficiency while maintaining model accuracy. DeepSeek-V3, the basic model of DeepSeek-R1, uses three mixed precisions: FP8, FP16, and FB32~\cite{liu2024deepseek,guo2025deepseek}. It combines the efficiency of FP8 and FP16 with the stability of FP32, strikes a balance between speed and accuracy, and improves reasoning efficiency while maintaining model accuracy.

\subsection{Chain-Of-Thought}

The Chain-Of-Thought is a method to guide large language models to perform step-by-step reasoning. It usually consists of three parts: instructions, logical basis, and examples. Compared with directly outputting answers, CoT allows the model to first perform logical decomposition when answering questions, and then get the final answer, thereby improving its reasoning ability and interpretability~\cite{wei2022chain}. DeepSeek-R1~\cite{guo2025deepseek} and o1-preview both use CoT technology, making them stronger in tasks such as logical reasoning, mathematical calculations, and language understanding. The chain of ideas improves reasoning ability in a variety of ways, such as enhancing interpretability, providing a step-by-step reasoning process, and making answers more transparent and credible~\cite{feng2023towards,xia2024beyond}. Through logical decomposition, the model is prevented from directly giving wrong answers, and the generalization ability is improved, which can be applied to problems in different fields, such as legal policy analysis, educational assessment, and language translation in the humanities and social sciences~\cite{karjus2025machine,thapa2025large}.

Most traditional Transformer language models generate text based on conditional probability, while models that use chain of ideas will force the model to generate intermediate steps first and then derive the final result during the reasoning process. Compared with the Transformer language model based on conditional probability, this method is closer to the human thinking mode and enables the model to perform more reliable reasoning. DeepSeek-R1 has a built-in CoT output format during the training phase, so when the user interacts with DeepSeek-R1, DeepSeek-R1 will automatically generate the reasoning process~\cite{guo2025deepseek}.

\subsection{Mixture Of Experts}
The core idea of Mixture of Experts is to divide the model into multiple experts, each of whom focuses on processing a specific type of data or task~\cite{cai2024survey,riquelme2021scaling,guo2025came}. Through the dynamic routing mechanism system, the most relevant experts are selected and activated according to the characteristics of the input data, thereby expanding the model capacity while maintaining computational efficiency, especially in large-scale tasks and multi-task learning.

MoE is usually composed of an expert network, a gating network, and a combination mechanism. The expert network is usually a set of parallel sub-networks. Each expert specializes in learning a specific pattern or feature of the input data~\cite{dimitri2025survey,li2025unraveling}. The gating network is an additional network, usually a softmax layer. Its role is to dynamically assign weights according to the input data and determine which experts should be activated. Through weight assignment, each input data can be processed by a different combination of experts instead of all experts participating in the calculation. The combination mechanism is that the weights output by the gating network are used to weightedly summarize the outputs of each expert to form the final prediction result.

Compared with the traditional model based on the standard Transformer architecture~\cite{vaswani2017attention},such as BERT~\cite{koroteev2021bert}, MoE has high computational efficiency and reduces computational cost~\cite{vats2024evolution}. Different experts can focus on different types of data, making the model more generalizable. MoE is suitable for large-scale tasks and has good performance in tasks such as natural language processing, such as in large language models such as DeepSeek-R1.

\newpage

\section{Experiments and analysis}

This section tests the performance of DeepSeek-R1 in seven areas of humanities and social sciences: low-resource language translation, educational question answering, student writing improvement in higher education, logical reasoning, educational measurement and psychometrics, public health policy analysis, and art education. The experimental results are compared with the results of o1-preview, which comes from \textit{Evaluation of OpenAI o1: Opportunities and Challenges of AGI}~\cite{zhong2024evaluation}(\url{https://arxiv.org/abs/2409.18486}).

\subsection{Low-Resource Language Translation}

In the evaluation of DeepSeek-R1 for low-resource language translation, we examine two Cherokee sentence cases and compare with the results of o1-preview(The test results of o1-preview come from \textit{Evaluation of OpenAI o1: Opportunities and Challenges of AGI}~\cite{zhong2024evaluation}). In this scenario, as shown in (Figure ~\ref{fig:Low-Resource_Language_Translation1}) below, the DeepSeek-R1 model translated the sentence as "Those children are playing soccer," while the o1-preview model produced "The boys are playing ball." Both models successfully identified all the key elements and provided detailed analyses of the sentence and its grammatical structure. However, it is worth noting that there are still some differences between DeepSeek-R1's "Those children" and "soccer" and o1-preview's "The boys" and "ball."

It can be observed that "The boys" is more precise. This is because "the boys" specifically refers to a group of male children, making the scope clear, whereas "Those children" has a broader scope and could include both boys and girls, making it less specific. If the context clearly refers to boys, using "The boys" is more accurate; if it refers to children in general, then "Those children" is more appropriate. Similarly, "soccer" is more precise. "Soccer" specifically refers to the sport of football, while "ball" is a more general term that could refer to any type of ball or spherical object. If the context clearly refers to football, using "soccer" is more accurate; if it refers to balls in general, then "ball" is more suitable.This example demonstrates the ability of both models to handle word recognition and basic grammar in low-resource language translation. However, while both models perform relatively well in dealing with unknown words and have their respective strengths, the choice between them depends on the specific context and the level of precision required.

\begin{figure}[H]
    \centering
    \includegraphics[width=1.0\textwidth]{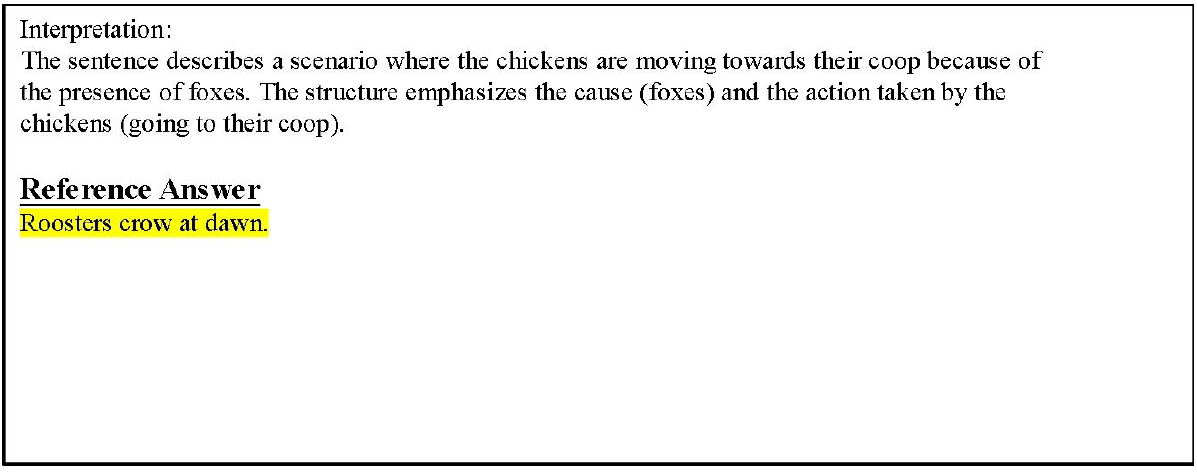} 
    \caption{\textbf{Low-Resource Language Translation: Case2.} Both DeepSeek-R1 and o1-preview face challenges in handling the subtle contextual details and precision required for more complex scenes.}
    \label{fig:Low-Resource_Language_Translation2}
\end{figure}

In the second scenario, as illustrated in (Figure ~\ref{fig:Low-Resource_Language_Translation2}), the DeepSeek-R1 model translated the sentence into "They are feeding the birds," whereas the o1-preview model rendered it as "The chickens are going to their coop because of the foxes." Unlike the first example, the two models provided entirely different translations for this sentence, both of which diverged significantly from the reference answer. The term "birds" in DeepSeek-R1's translation was not as precise as the "chickens" in o1-preview's version. While o1-preview accurately identified "chickens," it introduced additional interpretations, such as "foxes" and "coop." In contrast, DeepSeek-R1 did not include these extraneous associations. This example underscores that both models retain the capability to manage word recognition and basic grammar in low-resource language translation. However, they encounter challenges in handling the nuanced contextual subtleties and precision required in more complex scenarios.

\begin{figure}[H]
    \centering
    \includegraphics[width=1.0\linewidth, keepaspectratio]{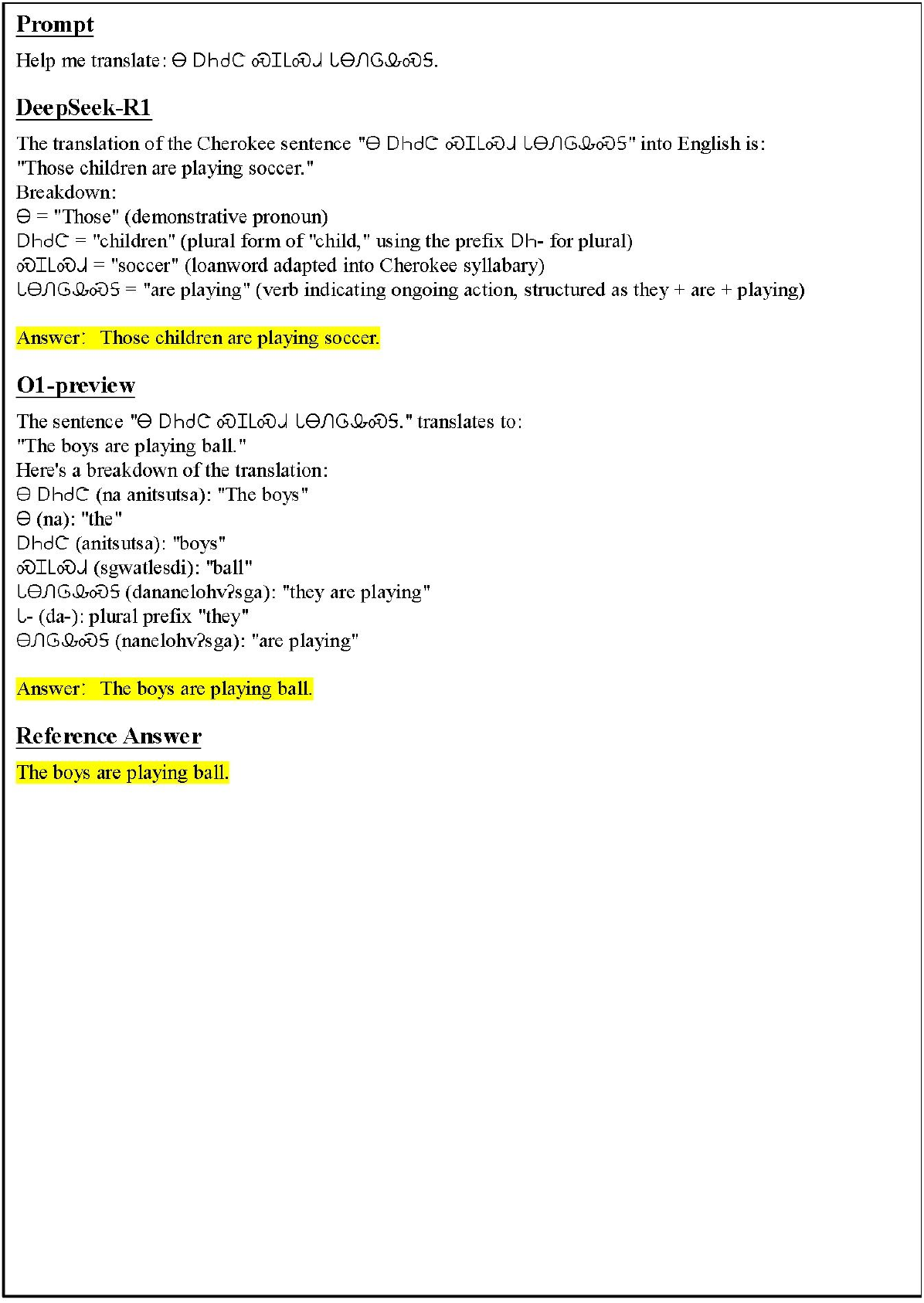}
    \caption{\textbf{Low-Resource Language Translation: Case1.} Both DeepSeek-R1 and o1-preview correctly translate the main information, but the details are slightly different.}
    \label{fig:Low-Resource_Language_Translation1}
\end{figure}

\begin{figure}[H]
    \centering
    \includegraphics[width=1.0\linewidth, keepaspectratio]{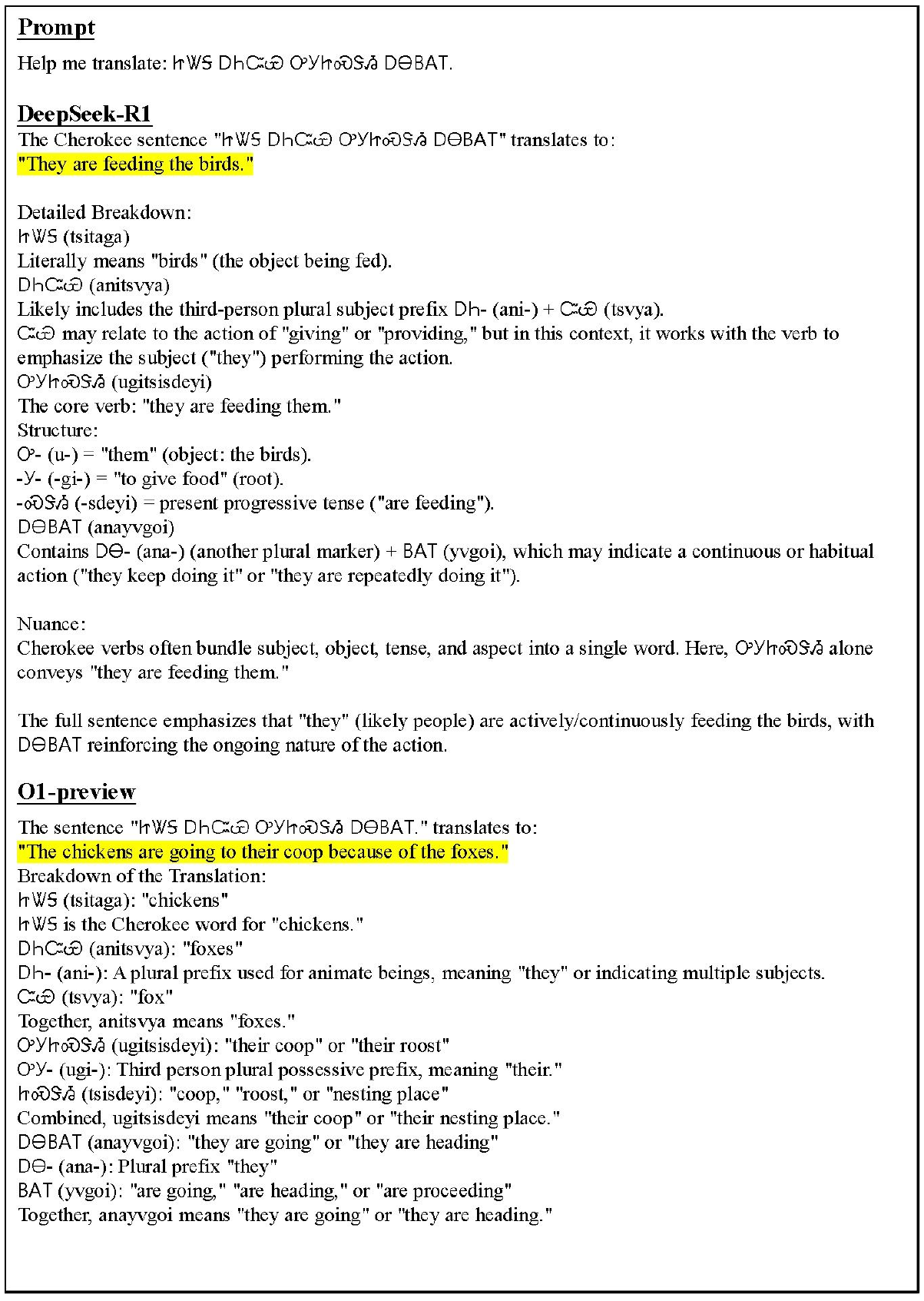}
\end{figure}
\newpage

\subsection{Student Writing Improvement in Higher Education}

\begin{figure}[H]
    \centering
    \includegraphics[width=1.0\textwidth]{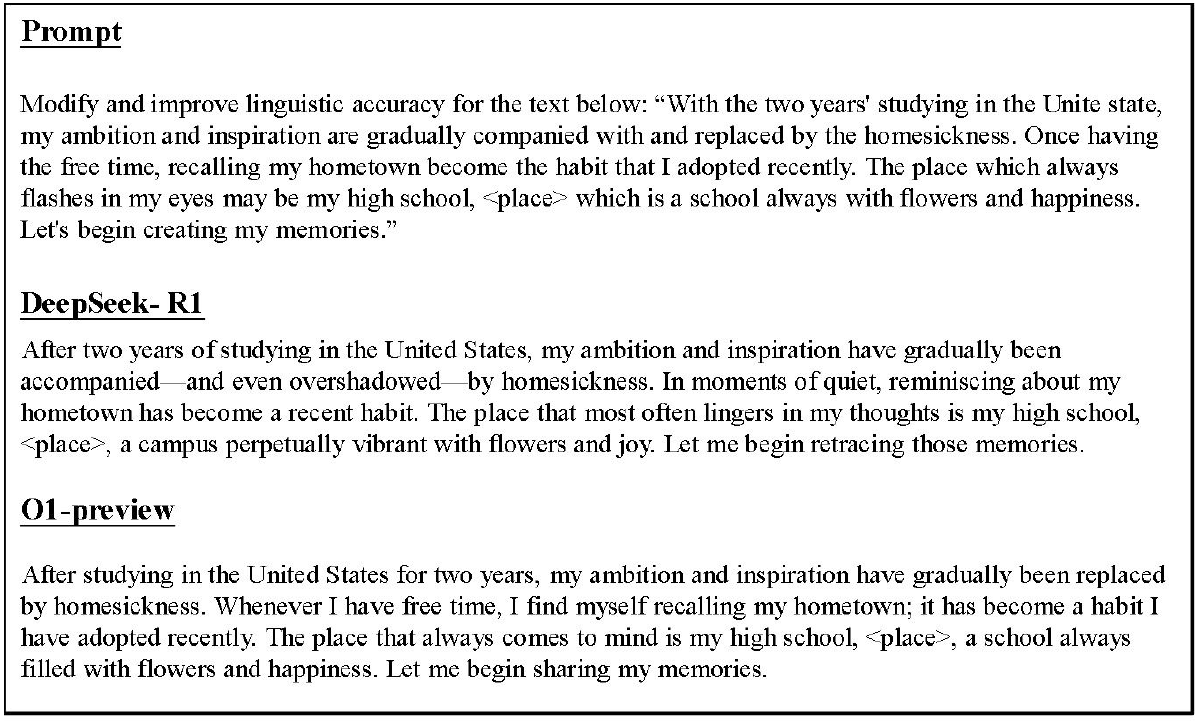} 
    \caption{\textbf{Student Writing Improvement in Higher Education:Case1.} Example of modifying and improving the language accuracy of the following text.}
    \label{fig:Student_Writing_Improvement_in_Higher_Education1}
\end{figure}

\begin{figure}[H]
    \centering
    \includegraphics[width=1.0\linewidth, keepaspectratio]{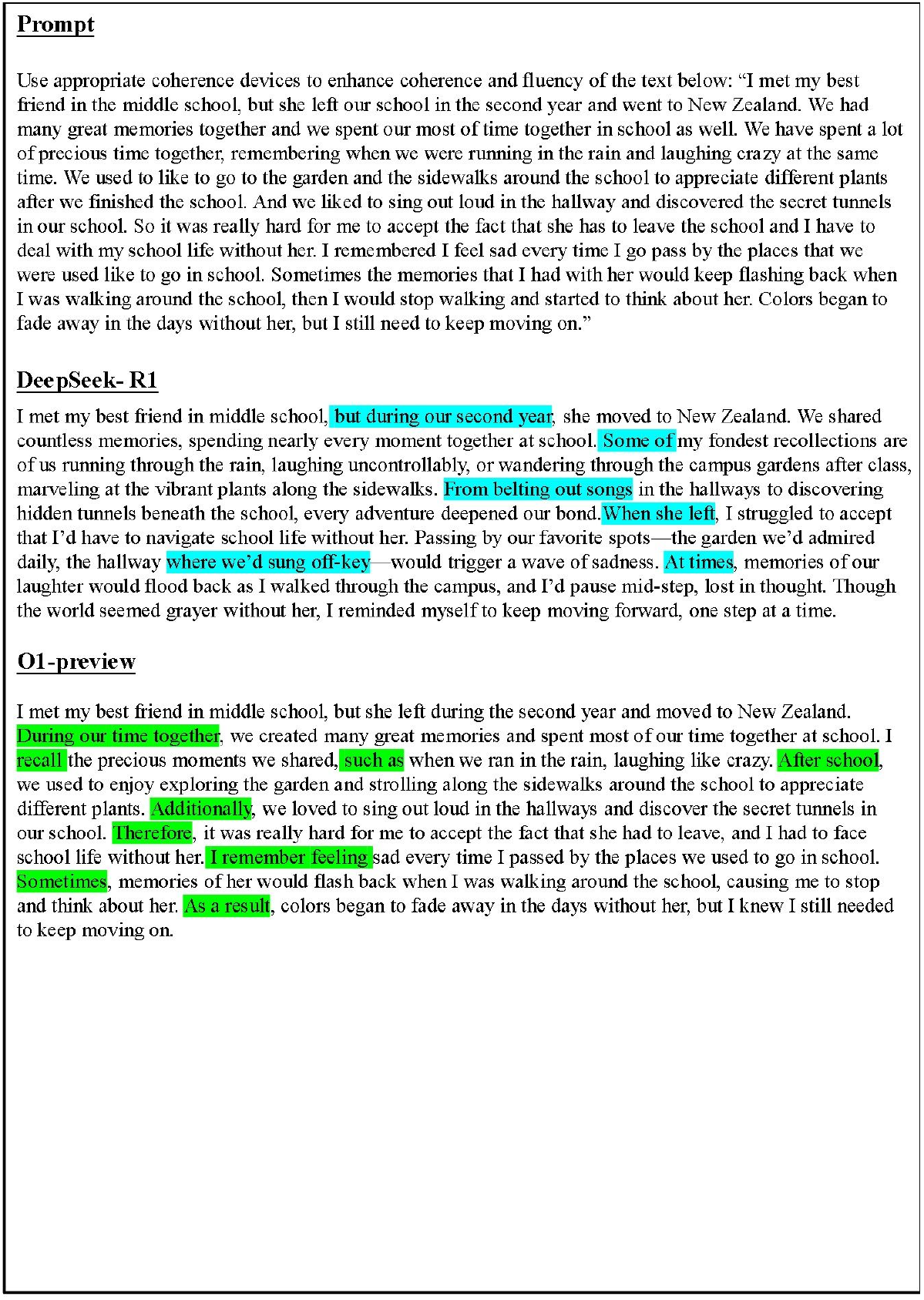}
    \caption{\textbf{Student Writing Improvement in Higher Education:Case2.} Example of appropriate coherence methods to enhance the coherence and flow of the following text}
    \label{fig:Student_Writing_Improvement_in_Higher_Education2}
\end{figure}

\begin{figure}[H]
    \centering
    \includegraphics[width=1.0\textwidth]{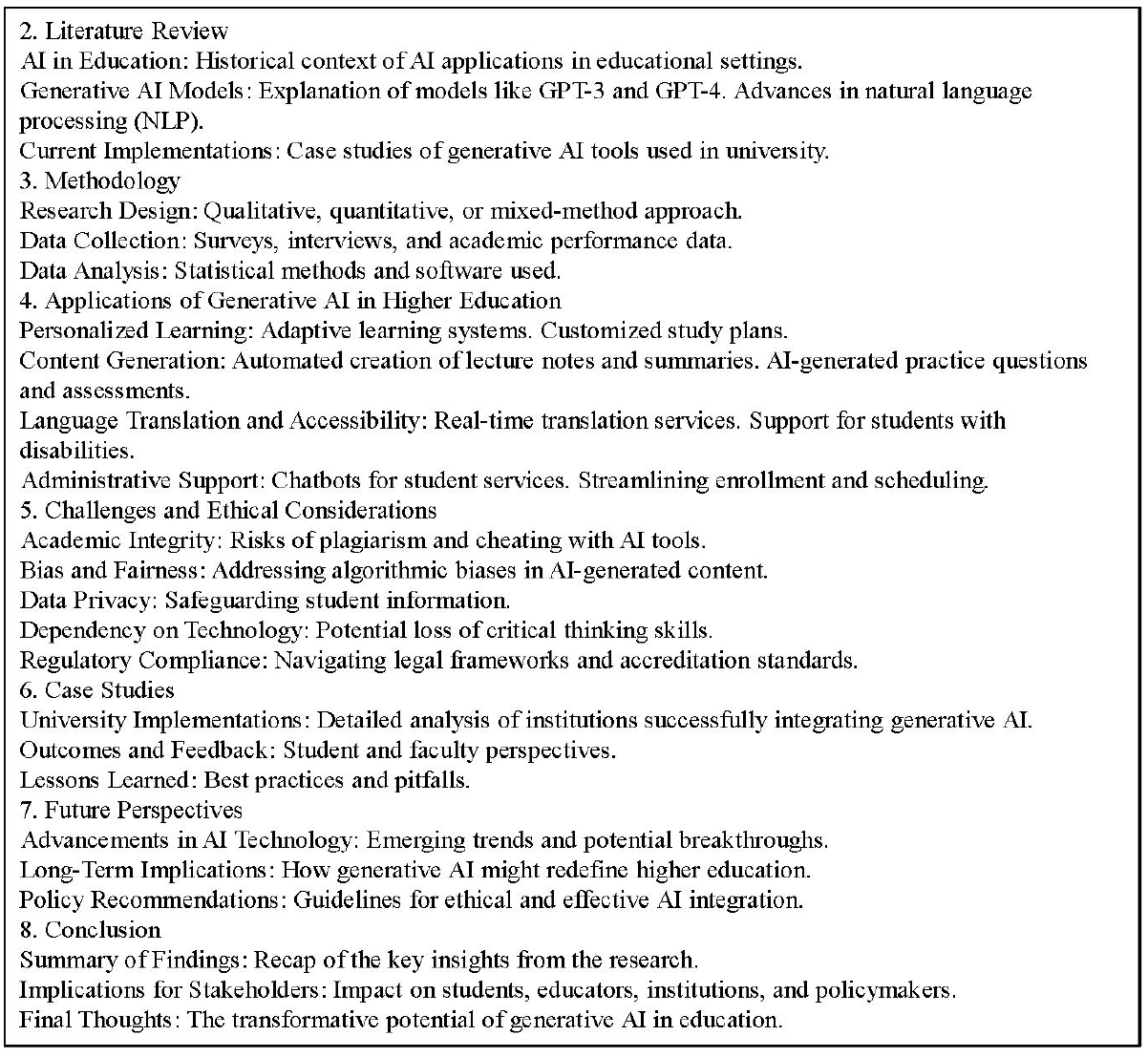} 
    \caption{\textbf{Student Writing Improvement in Higher Education:Case3.} Brief writing example given on the topic}
    \label{fig:Student_Writing_Improvement_in_Higher_Education3}
\end{figure}

\begin{figure}[H]
    \centering
    \includegraphics[width=1.0\linewidth, keepaspectratio]{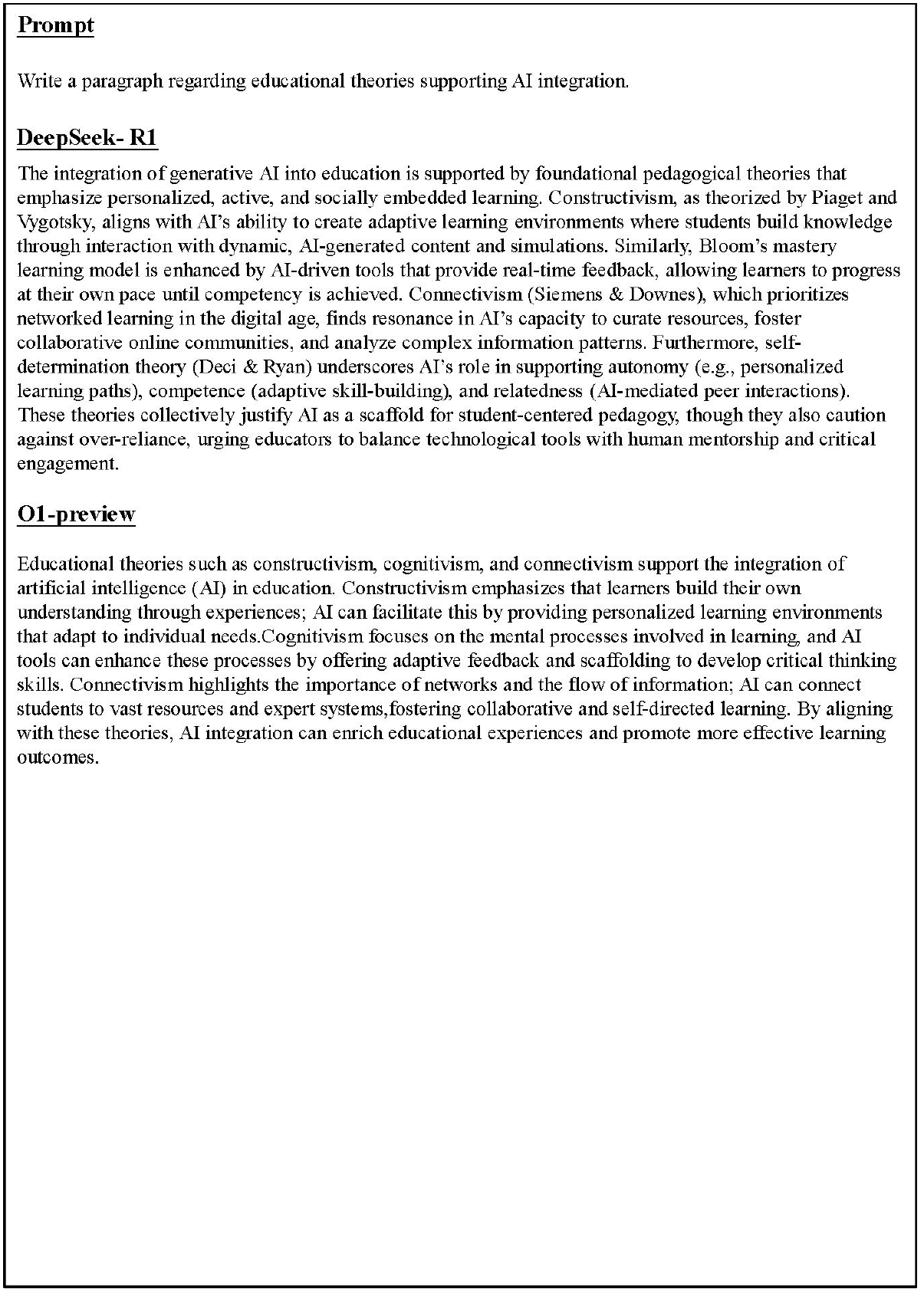}
    \caption{\textbf{Student Writing Improvement in Higher Education:Case4.} Example of paragraph writing on given topics}
    \label{fig:Student_Writing_Improvement_in_Higher_Education4}
\end{figure}

\begin{figure}[H]
    \centering
    \includegraphics[width=1.0\linewidth, keepaspectratio]{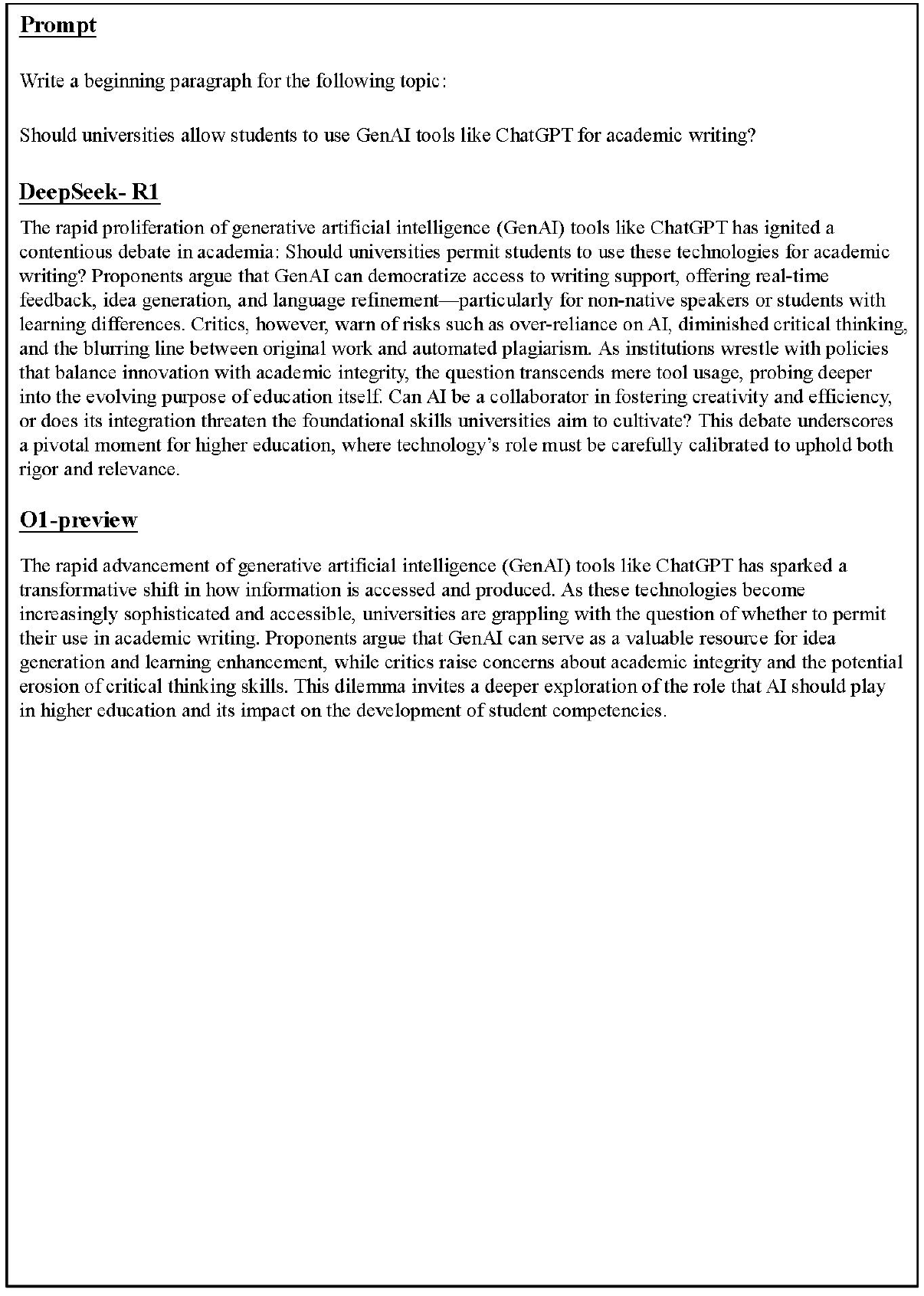}
    \caption{\textbf{Student Writing Improvement in Higher Education:Case5.} Example of writing the opening paragraph on a given topic}
    \label{fig:Student_Writing_Improvement_in_Higher_Education5}
\end{figure}

As educators continue to explore emerging technologies to enhance student learning, DeepSeek-R1 may serve as a pivotal pedagogical instrument in augmenting writing competencies among learners. A qualitative analysis of DeepSeek-R1's application in student writing (as illustrated in Figures \ref{fig:Student_Writing_Improvement_in_Higher_Education1},\ref{fig:Student_Writing_Improvement_in_Higher_Education2},\ref{fig:Student_Writing_Improvement_in_Higher_Education3},\ref{fig:Student_Writing_Improvement_in_Higher_Education4},\ref{fig:Student_Writing_Improvement_in_Higher_Education5}) has demonstrated promising outcomes across multiple dimensions. The comparative data reveal that while both DeepSeek-R1 and o1-preview models exhibit capacity to provide effective writing support—particularly in improving linguistic accuracy, enhancing textual coherence, and generating structural frameworks with preliminary conceptualization—significant divergences emerge in their output characteristics.

Notably, DeepSeek-R1 generates content demonstrates marked superiority in discursive depth, terminological precision, and affective articulation, frequently incorporating exemplification and citation practices that align with human cognitive patterns. In contrast, o1-preview outputs tend toward mechanistic theoretical analysis, characterized by limited referential grounding and reduced contextual adaptability(The test results of o1-preview come from \textit{Evaluation of OpenAI o1: Opportunities and Challenges of AGI}~\cite{zhong2024evaluation}).

However, within the domain of creative composition, while both systems exhibit positive pedagogical auxiliary value for writing skill development, their aggregate impact manifests substantial contextual variability contingent upon learners' specific requirements and implementation frameworks. This variability may precipitate passive adoption of preconfigured structural paradigms, potentially constraining writing efficacy and creative agency.

These observations necessitate critical reflection: although generative artificial intelligence (GenAI) constitutes a valuable supplementary resource, it cannot supplant the essential human-centric mentorship inherent in writing pedagogy. Crucially, current large language models (LLMs) remain developmental prototypes, demanding rigorous scrutiny of their limitations within educational ecosystems. We posit that subsequent research should prioritize longitudinal investigations into LLM functionalities and their pedagogical validity in tertiary education, with particular emphasis on developing evidence-based frameworks for optimizing their integration across diverse learning contexts.

\newpage

\begin{figure}[H]
    \centering
    \includegraphics[width=1.0\linewidth, keepaspectratio]{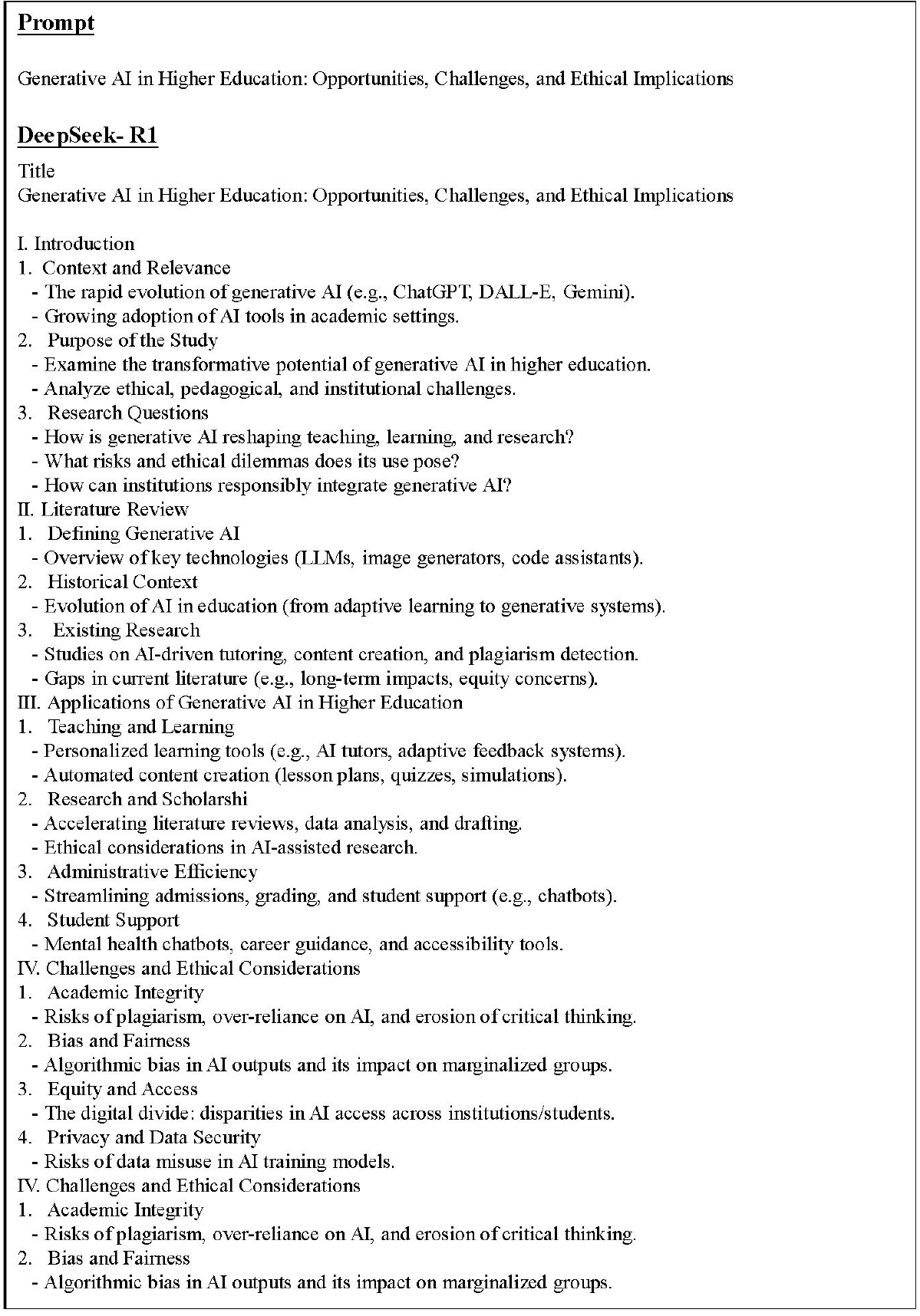}
\end{figure}
\begin{figure}[H]
    \centering
    \includegraphics[width=1.0\linewidth, keepaspectratio]{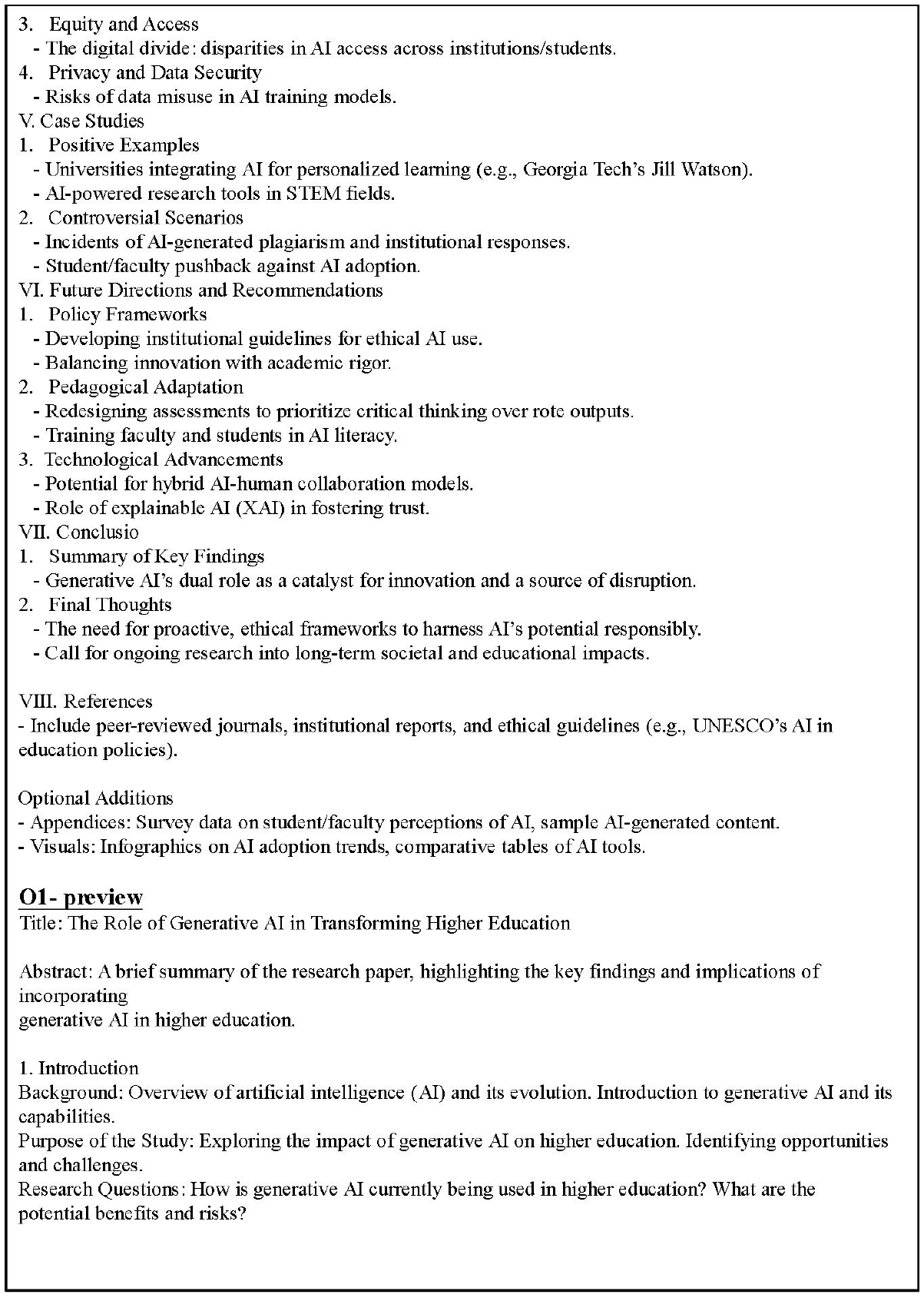}
\end{figure}

\subsection{Educational Q\&A}

In this section, we select geography and physics questions from the SciQ dataset to test DeepSeek-R1, and then compare the results with o1-preview(The test results of o1-preview come from \textit{Evaluation of OpenAI o1: Opportunities and Challenges of AGI}~\cite{zhong2024evaluation}). Both questions are choice questions. DeepSeek-R1 and o1-preview both give correct options, but they differ in analysis and answer structure.

\begin{figure}[H]
    \centering
    \includegraphics[width=1.0\linewidth, keepaspectratio]{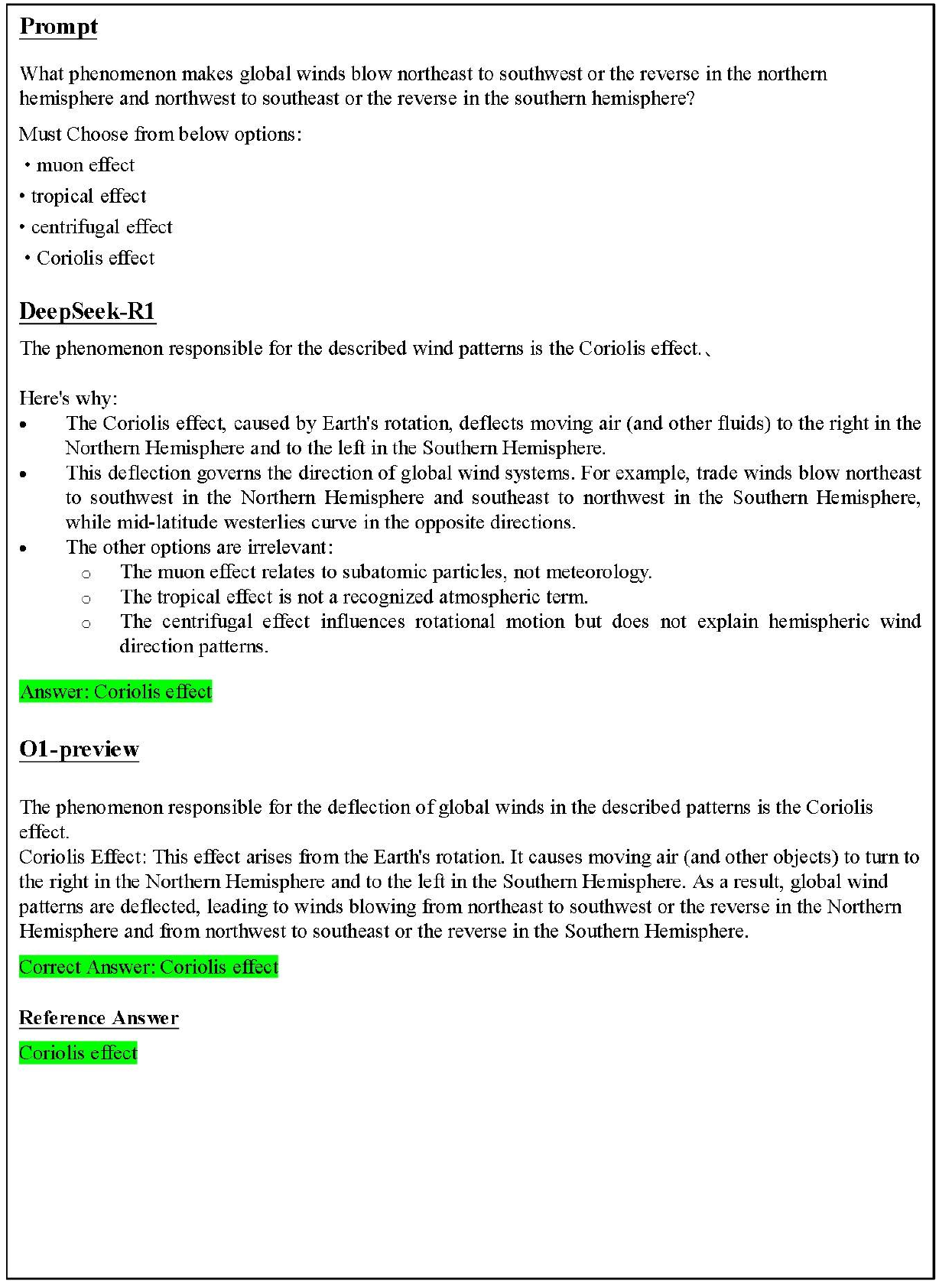}
    \caption{\textbf{Educational Q\&A:Case1.} Both DeepSeek-R1 and o1-preview gave the correct answer to the global wind deflection effect, but the analysis process was different. This question comes from the SciQ dataset.}
    \label{fig:Educational_ QA1}
\end{figure}

\begin{figure}[H]
    \centering
    \includegraphics[width=1.0\linewidth, keepaspectratio]{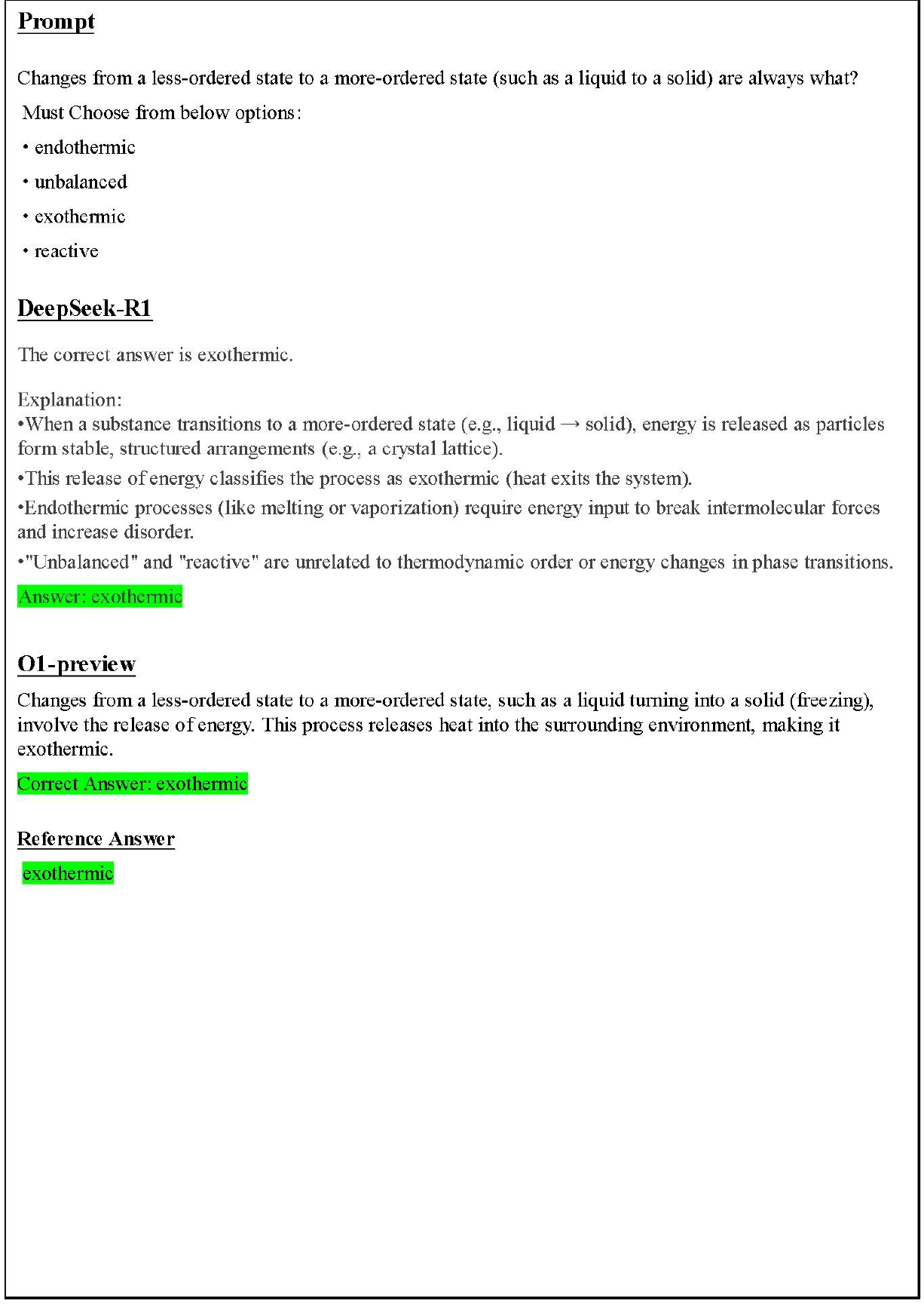}
    \caption{\textbf{Educational Q\&A:Case2.} Both DeepSeek-R1 and o1-preview gave the correct answer to the change process from disordered state to ordered state, but the analysis process was different. This question comes from the SciQ dataset.}
    \label{fig:Educational_ QA2}
\end{figure}

As shown in the (Figures \ref{fig:Educational_ QA1},\ref{fig:Educational_ QA2}),the results show that DeepSeek-R1 and o1-preview both gave correct answers to the two choice questions, but they differed in the analysis process and answer structure. Comparing the answers of the two, it can be found that DeepSeek-R1's answer is more detailed. It focuses on points and gives a structured answer. It first explains the selected answer at the beginning, and then gives the reasons for excluding the wrong options while analyzing the correct options. The answer is more similar to the reference answer in the textbook. O1-preview's answer focuses more on explaining the principles of the correct options, but does not analyze the wrong options. Its answer is presented in a coherent paragraph form, which is more compact overall and suitable for quick reading. For the field of education, if students want to explain learning and in-depth understanding in detail, the first answer is relatively better because it provides background information and excludes wrong options. If students need concise answers such as quick reference, the second answer is more suitable because it directly gives the answer and briefly explains the reason.

DeepSeek-R1 has performed well in the field of educational question answering, demonstrating the strong capabilities of large language models in the field of educational question answering and even in the humanities and social sciences. These models have a deep understanding of multidisciplinary concepts and can avoid misleading interference factors to choose the correct answer. DeepSeek-R1 and o1-preview each have their own focus in problem analysis. The analysis process of DeepSeek-R1 is similar to the reference answer and is more detailed, while the answer given by o1-preview focuses on the results and is suitable for quick reading. In short, large language models can improve efficiency in the field of education and provide personalized learning experiences, reduce the burden on teachers, and enhance students' autonomous learning ability. However, there are still some challenges in large language models, such as accuracy, interpretability, and in-depth understanding of teaching content, which are still directions for future improvements.

\newpage

\subsection{Logical Reasoning}

In the context of the accelerated development of artificial intelligence, logical reasoning ability has become an important indicator of the ability of large language models. DeepSeek-R1 has made significant progress in complex logical reasoning through a unique hybrid reasoning architecture,and optimization methods such as thought chaining and knowledge distillation. The experimental results show that DeepSeek-R1 has excellent reasoning ability and shows deep analytical ability in logical reasoning. Compared with o1-preview(The test results of o1-preview come from \textit{Evaluation of OpenAI o1: Opportunities and Challenges of AGI}~\cite{zhong2024evaluation}), DeepSeek-R1 has stronger divergent thinking, which may make DeepSeek-R1 give surprising answers to some questions, or may cause DeepSeek-R1 to answer some questions incorrectly.

\begin{figure}[H]
    \centering
    \includegraphics[width=1.0\textwidth]{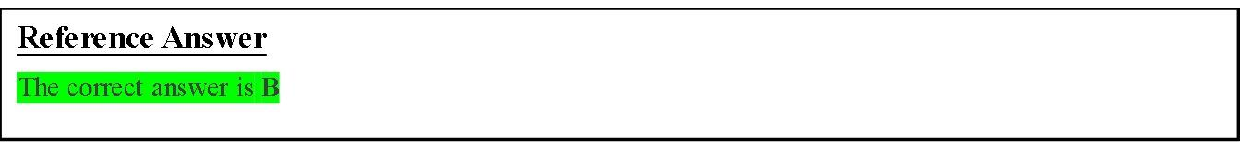} 
    \caption{\textbf{Logical Reasoning: Case1.} Both DeepSeek-R1 and o1-preview choose the correct answer for this question and give the detailed analysis process.}
    \label{fig:Logical_Reasoning1}
\end{figure}

As shown in the (Figure \ref{fig:Logical_Reasoning1}), both DeepSeek-R1 and o1-preview gave the correct answer and detailed analysis process to the question.From the perspective of answer format, DeepSeek-R1 first analyzes the four options separately, and explains which one may be the correct answer and which one is not during the analysis. Then two paragraphs are summarized and analyzed and the final conclusion is given. The answer of o1-preview is relatively concise. It first explains the correct option, then briefly explains the reasons for excluding other options, and finally summarizes and points out the correct option.In terms of the content of the answers, DeepSeek-R1 detailed the reasons for excluding the wrong options, while o1-preview simply briefly described that they were irrelevant to the question. In the analysis of the correct options, both mentioned "paid by the hour, if they can produce more products per hour, the labor cost per product will decrease, thus improving efficiency",The meaning of the analysis of the correct options is basically consistent with the reference answer.

\begin{figure}[H]
    \centering
    \includegraphics[width=1.0\textwidth]{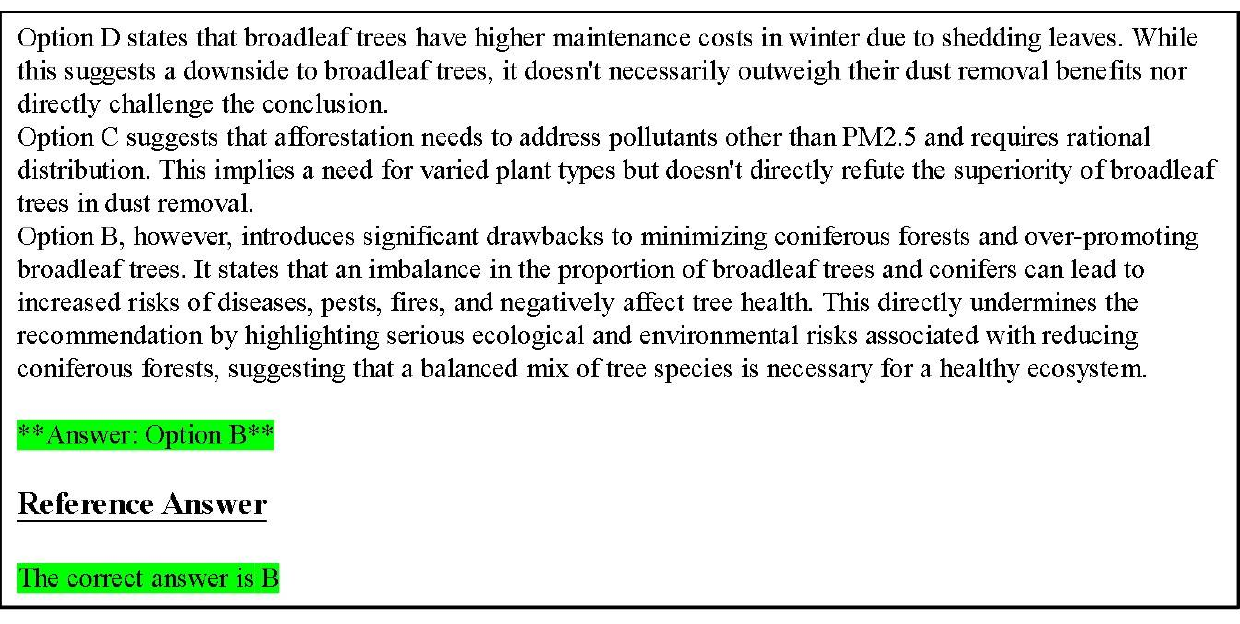} 
    \caption{\textbf{Logical Reasoning: Case2.} DeepSeek-R1 made the wrong choice for this question, while o1-preview gave the correct option.}
    \label{fig:Logical_Reasoning2}
\end{figure}

As shown in the (Figure \ref{fig:Logical_Reasoning2}),DeepSeek-R1 selected the wrong option D, while the answer given by o1-preview is the same as the reference answer, which is B.By analyzing the explanations given by DeepSeek-R1, we found that DeepSeek-R1 has strong divergent thinking, which may lead to wrong choices in logical reasoning due to excessive divergent thinking.The answer given by o1-preview is more in line with the general view and is the correct answer to the question. Therefore, in terms of logical reasoning, if the accuracy of the question is more important, o1-preview is more suitable.

The experimental results show that DeepSeek-R1 can still maintain high reasoning ability and high practicality when the training samples are limited. The analysis process of the logical reasoning module and the educational question-answering module DeepSeek-R1 and o1-preciew is similar. The analysis process given by DeepSeek-R1 is often more complex and detailed than o1-preview, which is suitable for in-depth understanding and learning, but it may also contain some wrong information. Therefore, users need to pay attention to identifying the content and refer to more authoritative information. The answer of o1-preview is suitable for quick reading and understanding. Both have their own strengths, and users can choose the model that suits their style for reference.Notably, DeepSeek-R1’s reasoning process exhibits innovative characteristics. In certain complex cases, its conclusions may diverge from reference answers(As shown in the (Figure \ref{fig:Logical_Reasoning2})).

\begin{figure}[H]
    \centering
    \includegraphics[width=1.0\linewidth, keepaspectratio]{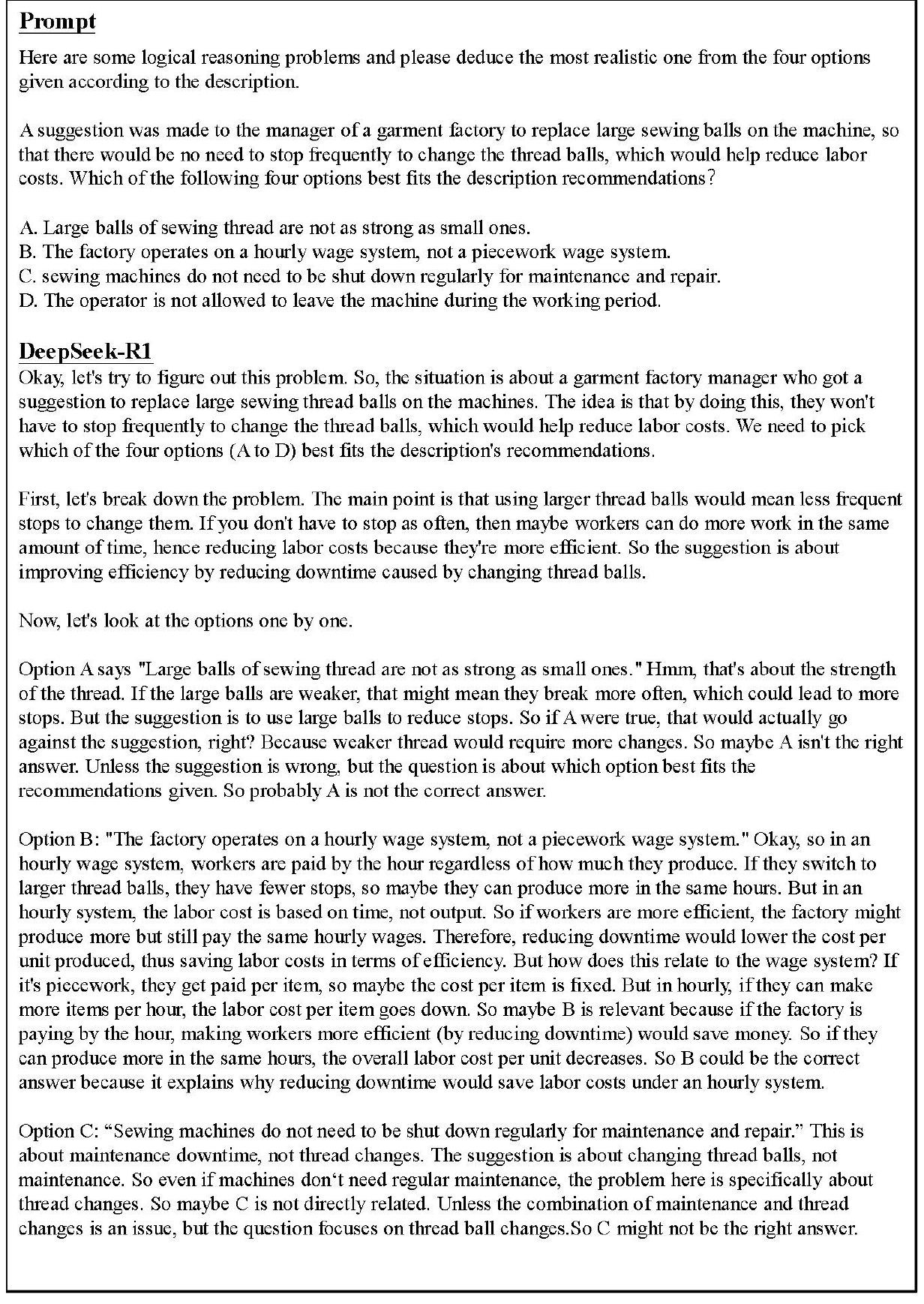}
\end{figure}
\begin{figure}[H]
    \centering
    \includegraphics[width=1.0\linewidth, keepaspectratio]{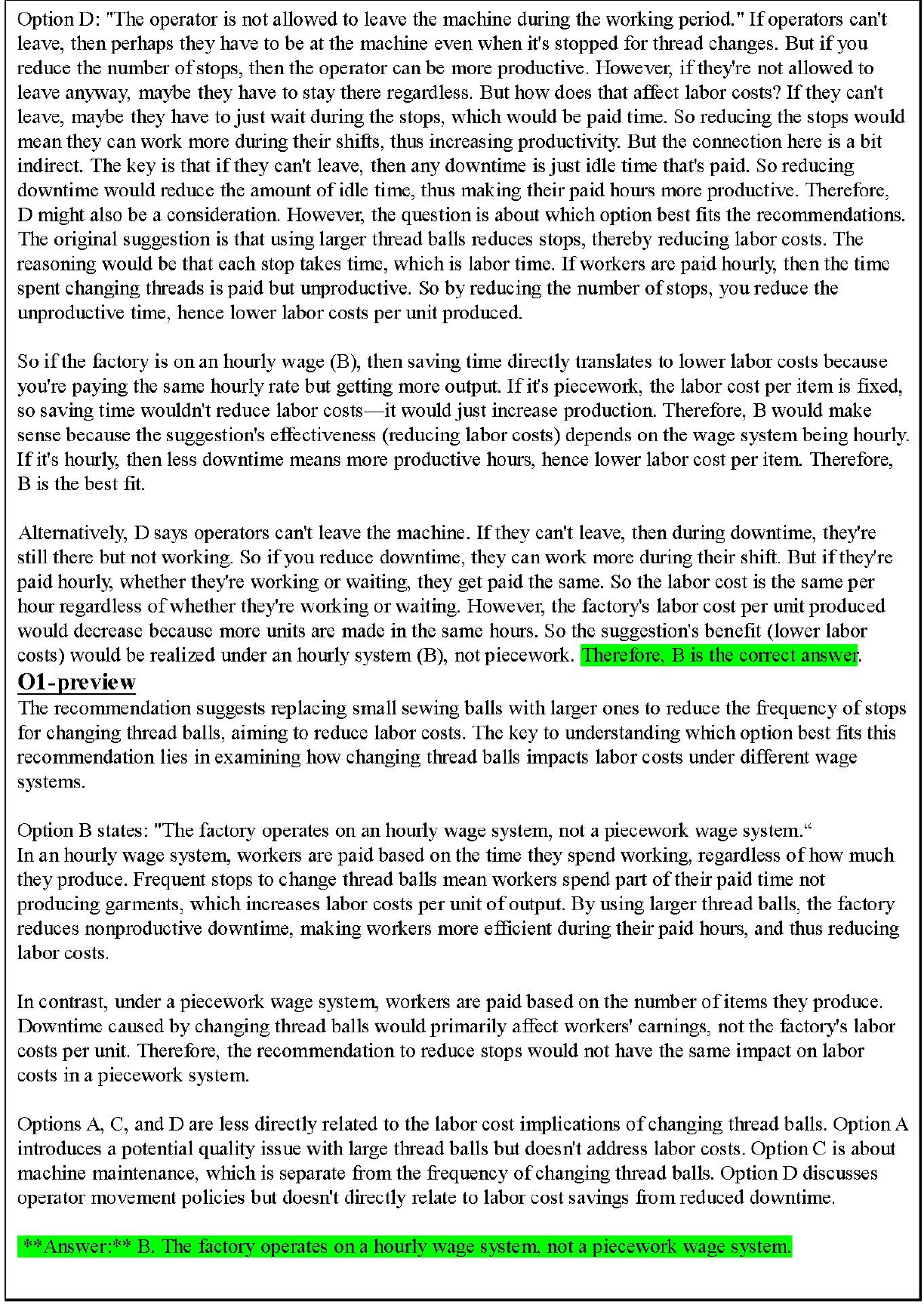}
\end{figure}

\begin{figure}[H]
    \centering
    \includegraphics[width=1.0\linewidth, keepaspectratio]{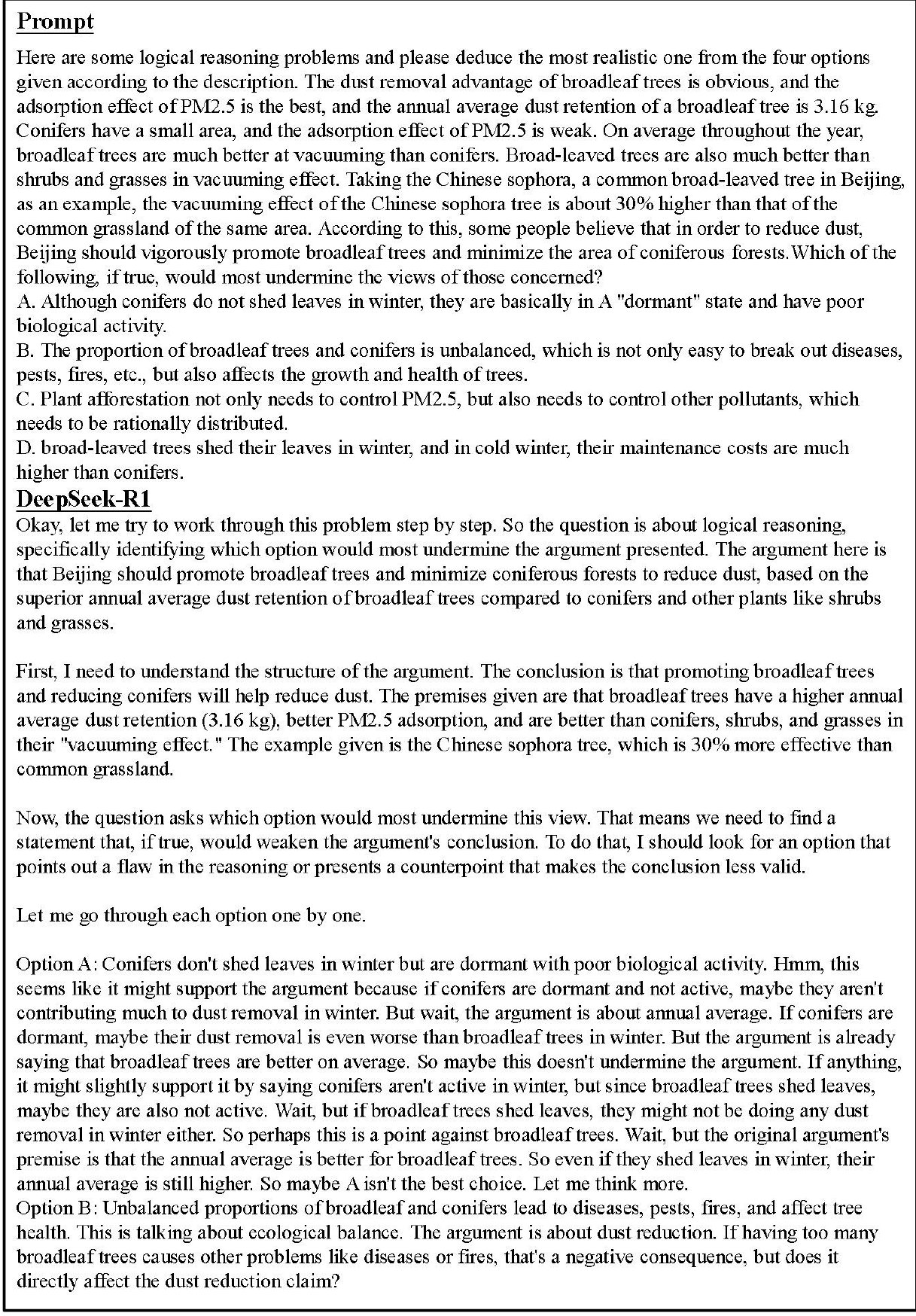}
\end{figure}
\begin{figure}[H]
    \centering
    \includegraphics[width=1.0\linewidth, keepaspectratio]{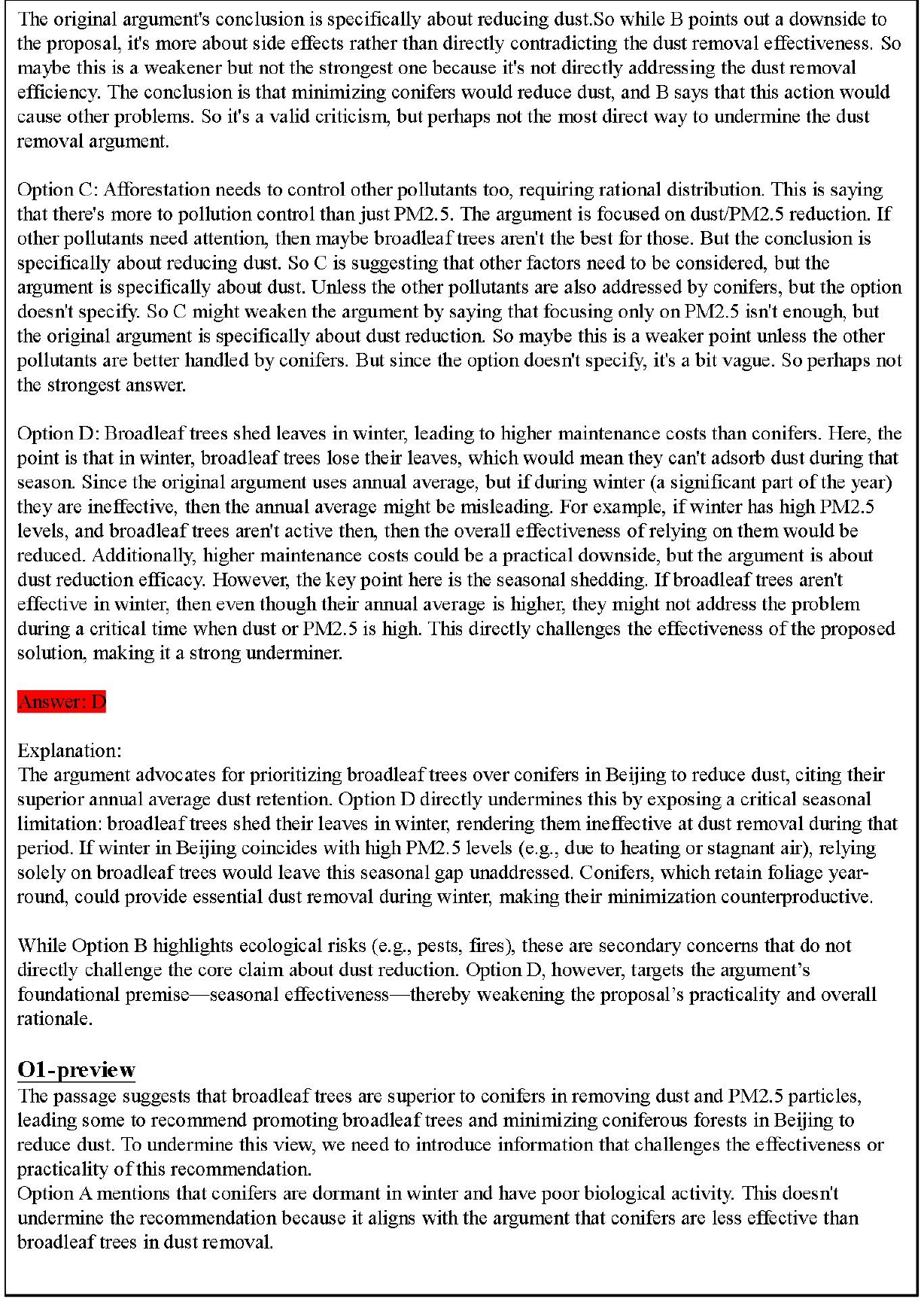}
\end{figure}

\newpage

\subsection{Educational Measurement and Psychometrics}

This section aims to test the performance of DeepSeek-R1 in analyzing complex topics in educational measurement and psychometrics, and compare it with the results of o1-preview(The test results of o1-preview come from \textit{Evaluation of OpenAI o1: Opportunities and Challenges of AGI}~\cite{zhong2024evaluation}). By evaluating their responses to three specific cases, we assess their depth, accuracy, and comprehensiveness in addressing key psychometric concepts.  
  
\begin{figure}[H]
    \centering
    \includegraphics[width=1.0\linewidth, keepaspectratio]{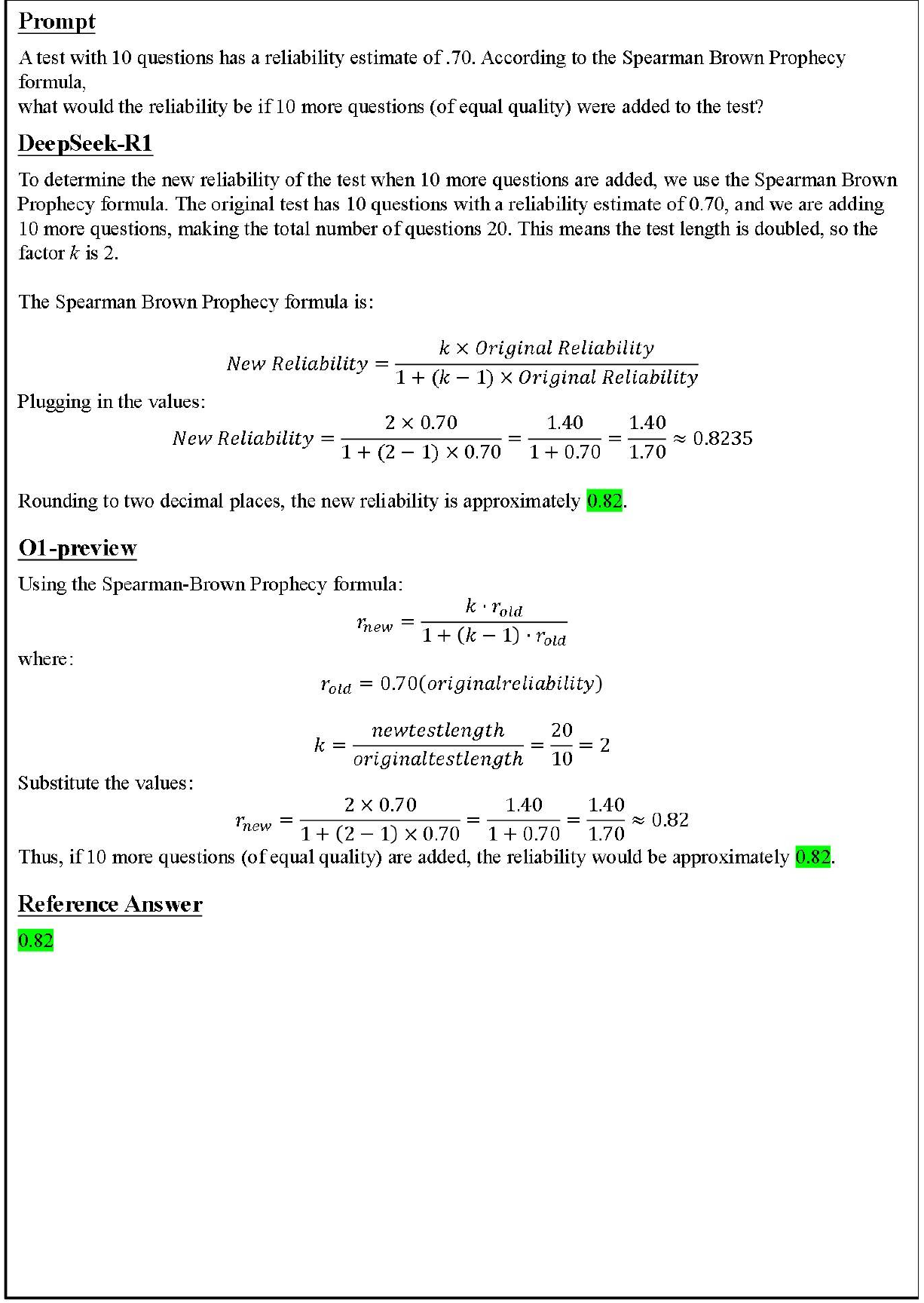}
    \caption{\textbf{Educational Measurement and Psychometrics: Case 1.} An Example Demonstrating the Ability of DeepSeek-R1 and O1-preview to Perform Psychometric Analysis Using the Spearman-Brown Prophecy Formula.}
    \label{fig:Educational_Measurement_and_Psychometrics1}
\end{figure}

Case 1 (Figure~\ref{fig:Educational_Measurement_and_Psychometrics1}) examines the use of the Spearman-Brown prophecy formula to predict reliability changes when adding 10 more test items of equal quality. Both models correctly estimated the new reliability to be approximately 0.82 but adopted different approaches. DeepSeek-R1 provided a more detailed explanation, explicitly defining key variables such as \( k \) (test length multiplier) and the initial reliability. It systematically explained each step, highlighting the logical reasoning behind the computation and clarifying the implications of test length changes on reliability. In contrast, o1-preview prioritized efficiency, directly substituting values into the formula to arrive at the result quickly. This suggests that DeepSeek-R1 excels in providing comprehensive explanations, which enhances conceptual understanding.  

\begin{figure}[H]
    \centering
    \includegraphics[width=1.0\linewidth, keepaspectratio]{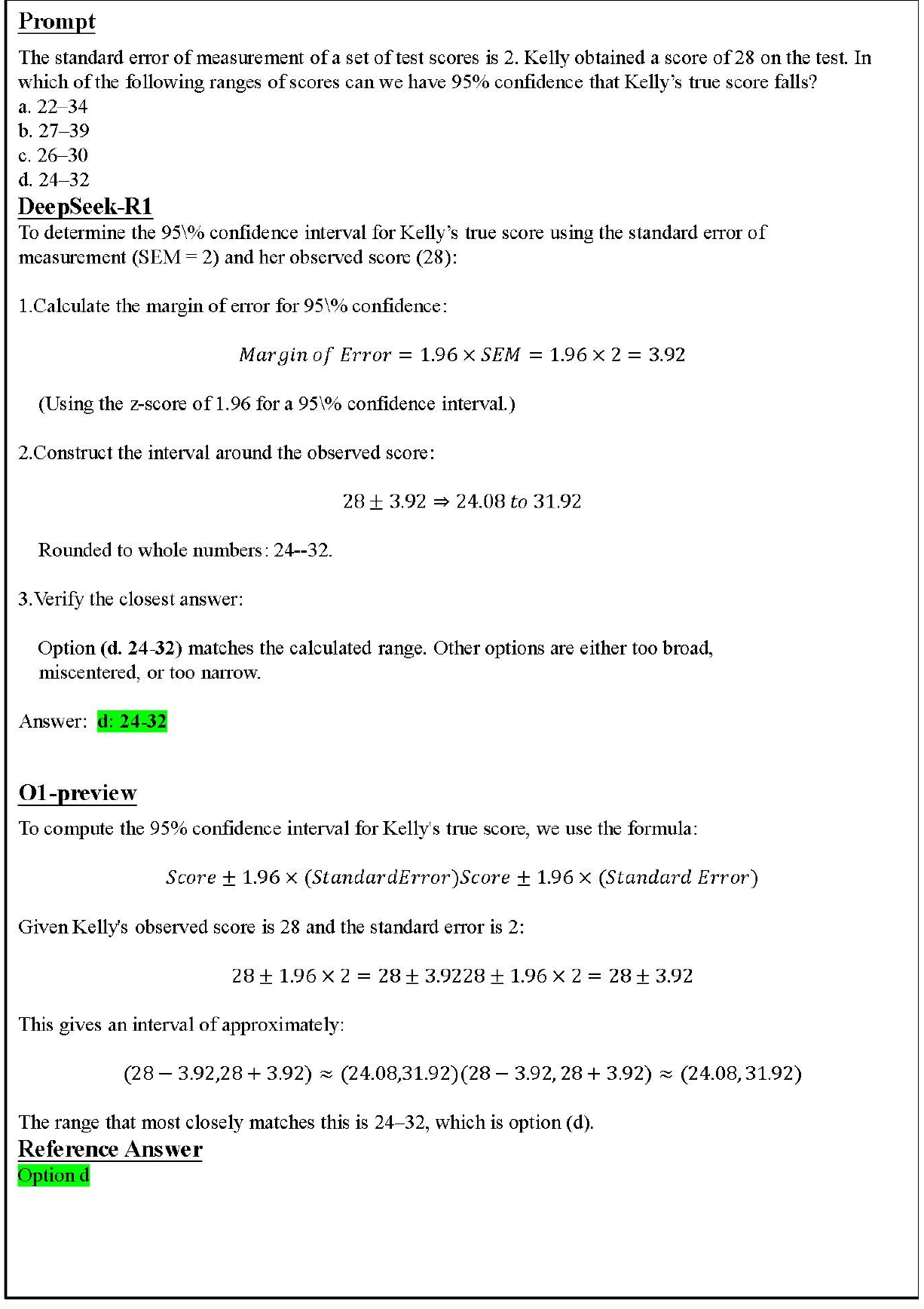}
    \caption{\textbf{Educational Measurement and Psychometrics: Case 2.} An Example Illustrating the Capability of DeepSeek-R1 and O1-preview to Perform Psychometric Analysis Using the Standard Error of Measurement. Green Indicates Consistency Between the LLMs’ Response and the Reference Answer.}
    \label{fig:Educational_Measurement_and_Psychometrics2}
\end{figure}

Case 2 (Figure~\ref{fig:Educational_Measurement_and_Psychometrics2}) explores the calculation of Kelly’s true score confidence interval at the 95\% level using a standard error of measurement (SEM) of 2 and an observed score of 28. Both models correctly identified the confidence interval as 24–32 (answer choice d), but their analytical approaches differed. DeepSeek-R1 broke down the calculation in detail, explaining the multiplication of 1.96 by the SEM (1.96 × 2 = 3.92), rounding considerations, and the justification for selecting option (d). Additionally, it analyzed why alternative answer choices were incorrect, improving the interpretability of the solution. In contrast, o1-preview performed the computation efficiently but without extensive elaboration. The structured reasoning provided by DeepSeek-R1 makes it particularly valuable in educational settings where a deeper understanding of statistical logic is essential.  

\begin{figure}[H]
    \centering
    \includegraphics[width=1.0\linewidth, keepaspectratio]{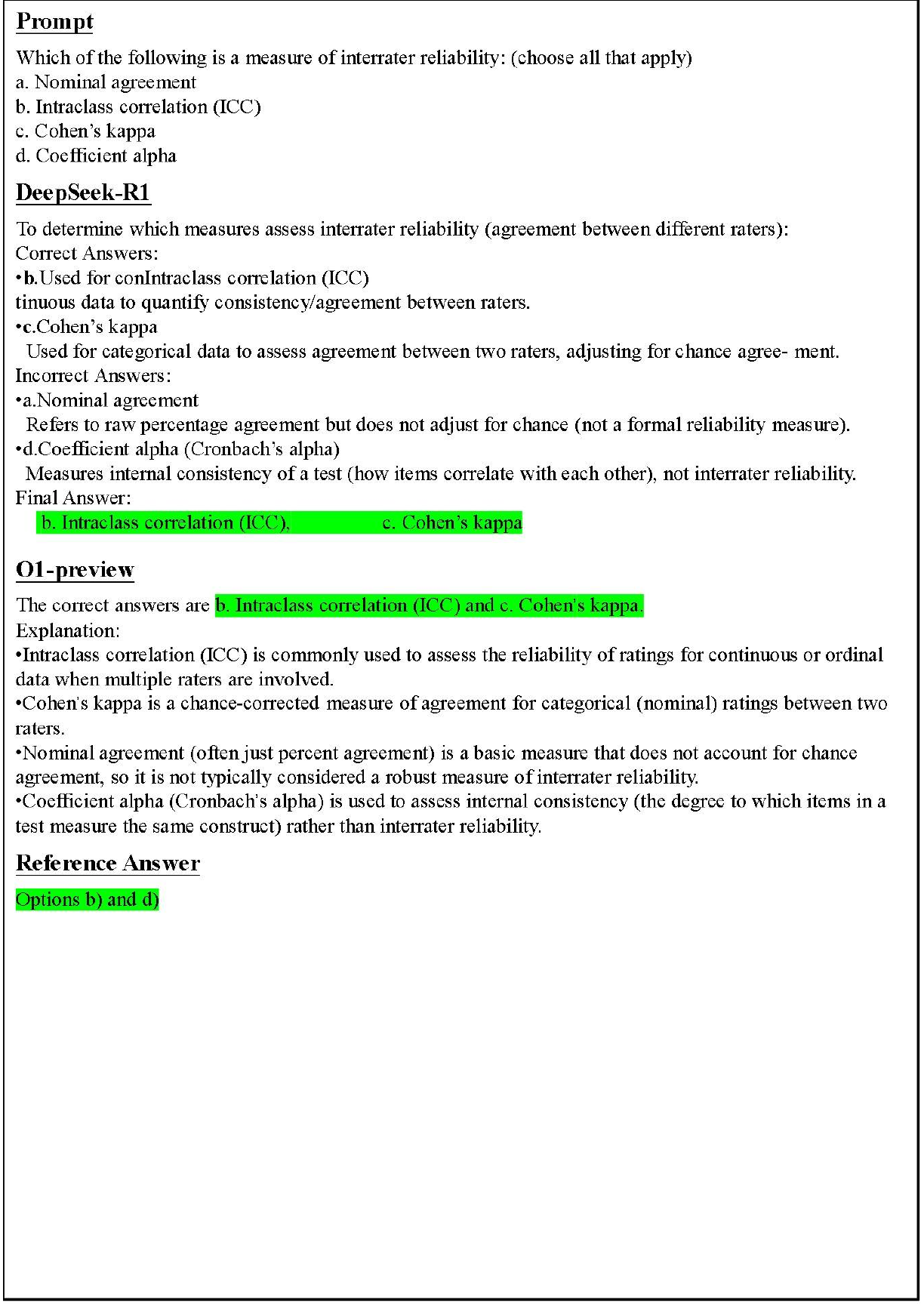}
    \caption{\textbf{Educational Measurement and Psychometrics: Case 3.} An Example Illustrating the Capability of DeepSeek-R1 and O1-preview to Assess Interrater Reliability Measures. Green Indicates Consistency Between the LLMs’ Response and the Reference Answer.}
    \label{fig:Educational_Measurement_and_Psychometrics3}
\end{figure}

Case 3 (Figure~\ref{fig:Educational_Measurement_and_Psychometrics3}) evaluates different statistical methods for assessing interrater reliability. Both models correctly identified (b) Intraclass Correlation Coefficient (ICC) and (c) Cohen’s kappa as suitable methods. However, DeepSeek-R1 provided a more in-depth discussion. It not only confirmed that ICC measures agreement among raters for continuous data but also elaborated on its role in quantifying rater consistency. Moreover, DeepSeek-R1 explicitly stated why nominal agreement is not considered a valid reliability metric and provided a detailed analysis of why Cronbach’s alpha applies to internal consistency rather than interrater reliability. While O1-preview offered a concise response, DeepSeek-R1's structured explanation of incorrect options enhances the reader’s ability to distinguish between interrater reliability and internal consistency.  

DeepSeek-R1, in contrast, provided a more detailed discussion. It not only confirmed that ICC measures agreement among raters using continuous data but also elaborated on its role in quantifying rater consistency. Furthermore, it explicitly stated why nominal agreement is not considered a valid reliability metric and provided a more in-depth analysis of why Cronbach’s alpha applies to internal consistency rather than interrater reliability. By offering a more structured explanation of incorrect options, DeepSeek-R1 enhances the reader’s understanding of the distinctions between interrater reliability and internal consistency.  

Despite the limited scope of this evaluation, the results highlight significant differences in how o1-preview and DeepSeek-R1 handle psychometric analysis. Overall, DeepSeek-R1 demonstrates superior depth in explanation, logical reasoning, and contextual understanding, making it particularly valuable for educational settings requiring conceptual clarity. O1-preview, while also providing accurate responses, excels in concise and efficient computations.

\newpage

\subsection{Public Health Policy Analysis}

The purpose of this section is to demonstrate the ability of DeepSeek-R1 to analyze public health policy issues by its performance in answering questions about the Affordable Care Act (ACA), then compare with the results of o1-preview(The test results of o1-preview come from \textit{Evaluation of OpenAI o1: Opportunities and Challenges of AGI}~\cite{zhong2024evaluation}).We will evaluate the depth, accuracy, and comprehensiveness of these two models in detail based on three specific questions and their reference answers.

\begin{figure}[H]
    \centering
    \includegraphics[width=1.0\linewidth, keepaspectratio]{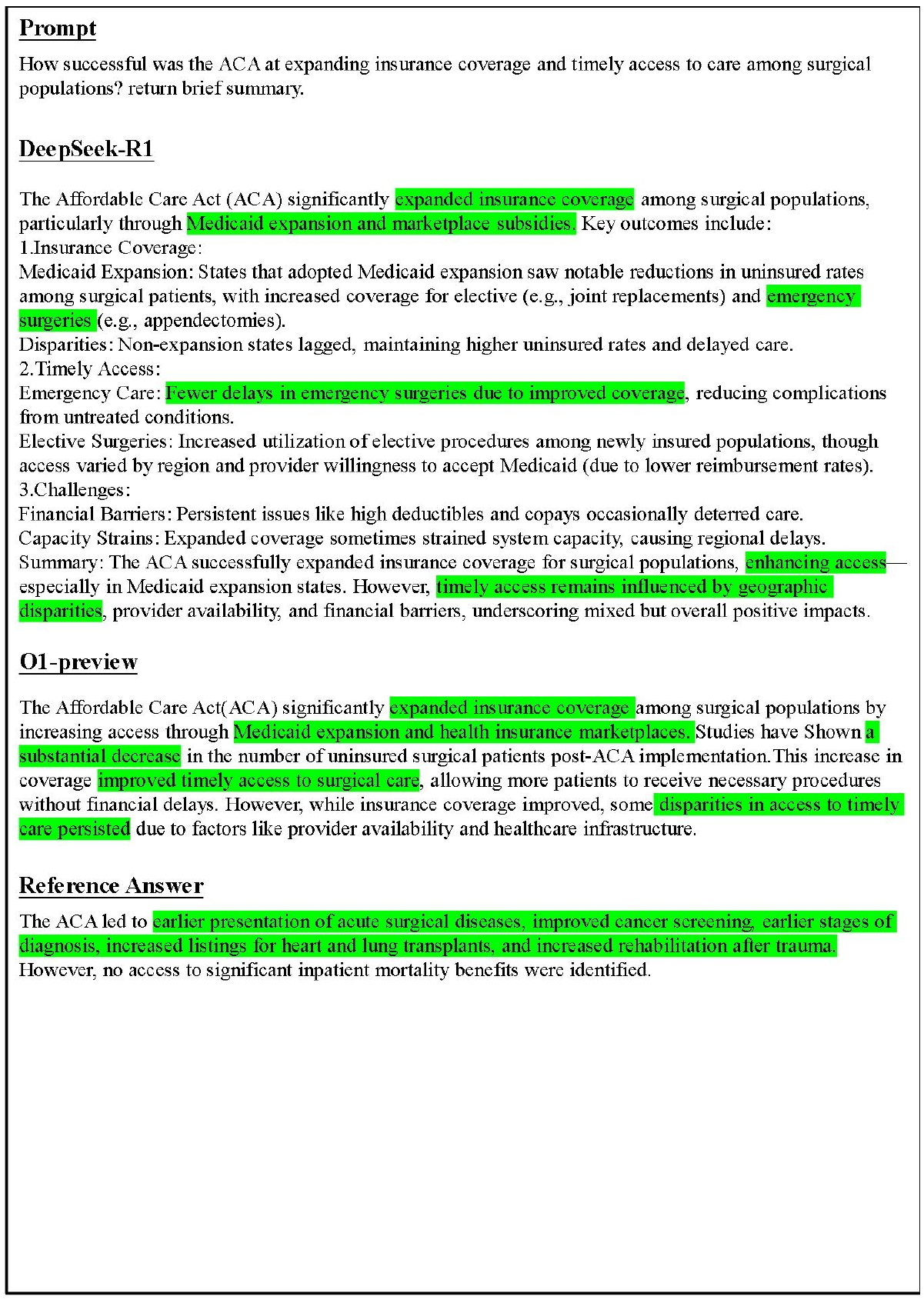}
    \caption{\textbf{Public Health Policy Analysis: Case 1.} one example illustrating the capabilities of Deepseek-R1 and o1-preview to respond to specialized questions of health policy in the domain of Public Health. Green indicates consistency between the LLMs' response and the reference answer.}
    \label{fig:Public_Health_Policy_Analysis1}
\end{figure}

\begin{figure}[H]
    \centering
    \includegraphics[width=1.0\textwidth]{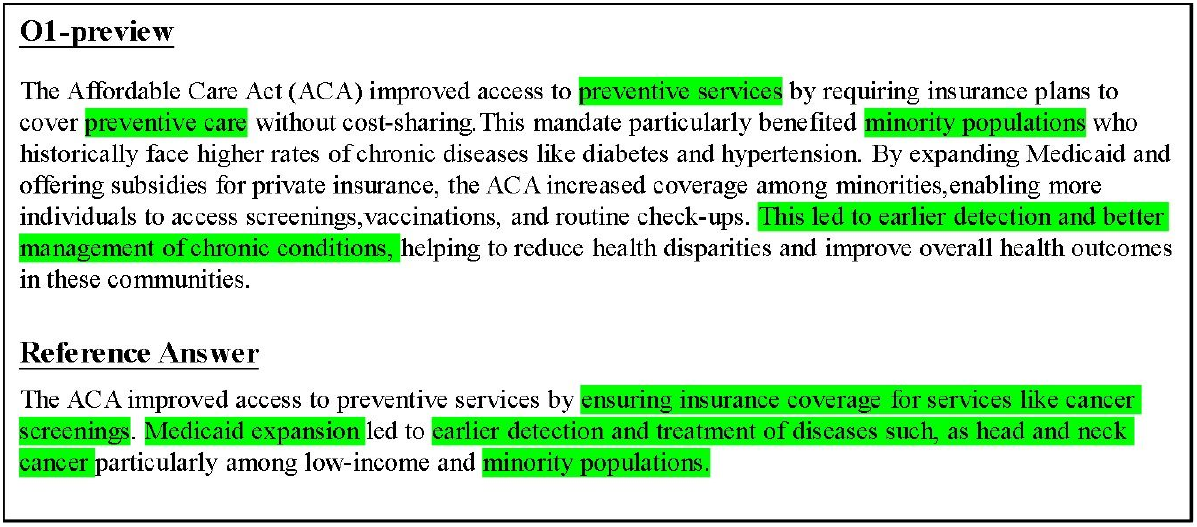} 
    \caption{\textbf{Public Health Policy Analysis: Case 2.} one example illustrating the capabilities of Deepseek-R1 and o1-preview to respond to specialized questions of health policy in the domain of Public Health. Green indicates consistency between the LLMs' response and the reference answer.}
    \label{fig:Public_Health_Policy_Analysis2}
\end{figure}

\begin{figure}[H]
    \centering
    \includegraphics[width=1.0\textwidth]{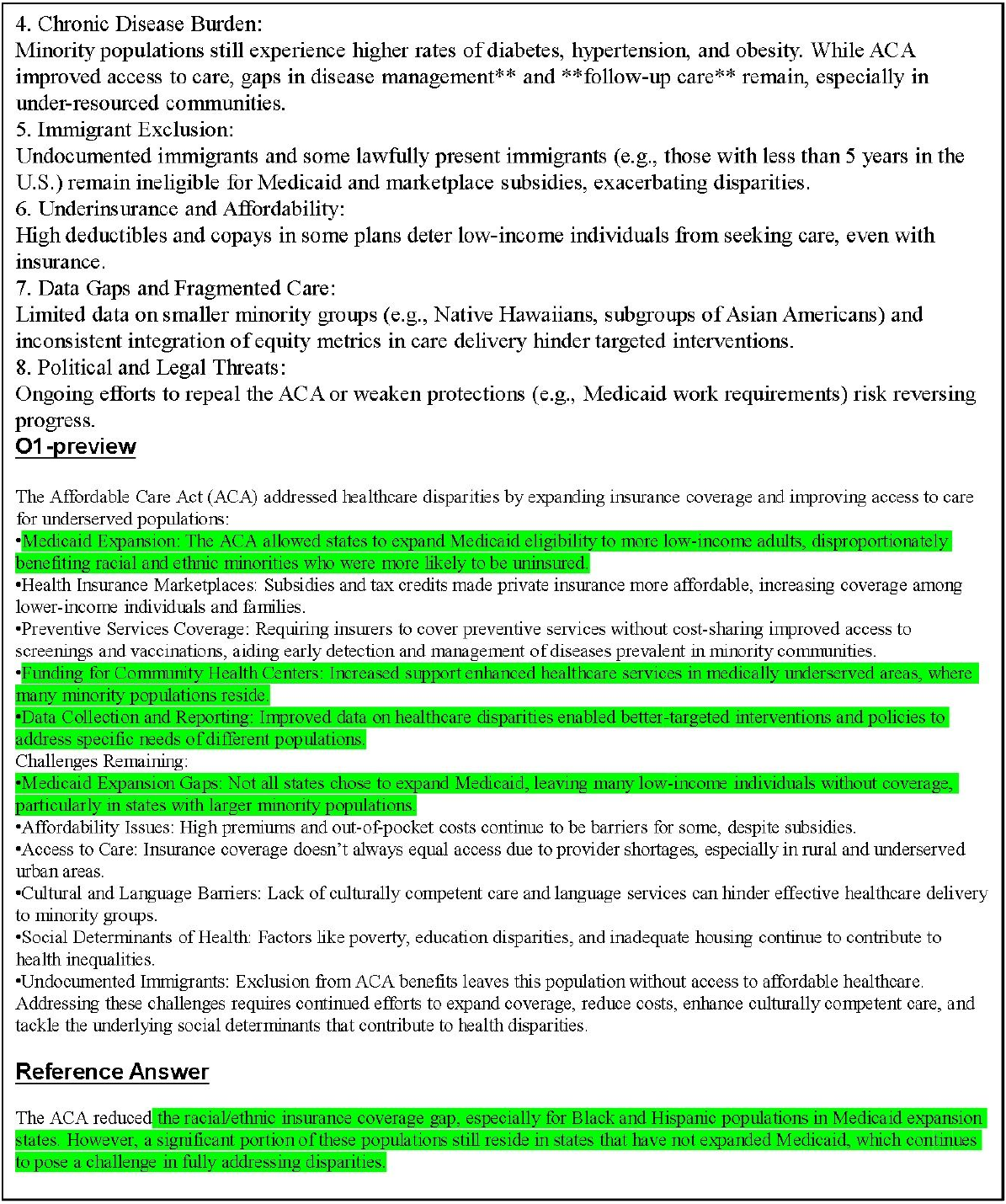} 
    \caption{\textbf{Public Health Policy Analysis: Case 3.} one example illustrating the capabilities of Deepseek-R1 and o1-preview to respond to specialized questions of health policy in the domain of Public Health. Green indicates consistency between the LLMs' response and the reference answer.}
    \label{fig:Public_Health_Policy_Analysis3}
\end{figure}

The case 1 (Figure~\ref{fig:Public_Health_Policy_Analysis1}) explores the impact of the Affordable Care Act in expanding insurance coverage and improving timely access to care for surgical patients. DeepSeek-R1 provides a detailed analysis that not only discusses changes in coverage for different types of surgeries but also identifies challenges such as financial barriers and system capacity constraints.The case 2 (Figure~\ref{fig:Public_Health_Policy_Analysis2}) focuses on the impact of the ACA on preventive services, particularly the role of chronic disease management for minorities. DeepSeek-R1 delves into a variety of mechanisms, including free cost-sharing coverage of preventive services, expansion of the Medicaid program, support of community-based programs, and protections from discrimination for people with pre-existing conditions. In addition, DeepSeek-R1 discusses the impact of social determinants on health outcomes, demonstrating a greater understanding of social context.The case 3 (Figure~\ref{fig:Public_Health_Policy_Analysis3}) discusses how the ACA is addressing health care inequities and the challenges it faces. DeepSeek-R1 outlines the ACA's key initiatives, such as expanding Medicaid coverage and supporting community health centers, but it also identifies some persistent challenges. And it uses specific data to support its points, with a particular focus on structural issues such as racism and bias. This makes DeepSeek-R1's responses more comprehensive and compelling. In contrast, o1-preview's answers were more general, focusing on general trends such as Medicaid expansion and reductions in the number of uninsured patients, but lacked specific analysis of mechanisms.

Despite the limited size of the current dataset, this evaluation revealed significant differences between DeepSeek-R1 and o1-preview in public health policy analysis. Overall, DeepSeek-R1 excels in providing detailed analysis, data support, and structured discussion, and is particularly well suited for tasks that require deeper understanding and integration of information from multiple sources. In contrast, o1-preview, while also providing valuable information, is slightly less detailed and comprehensive. This comparison not only helps to understand the role of existing models in public health decision support but also points the way for future research. In future evaluations, we may consider using larger datasets, such as the Compilation of State Data on the Affordable Care Act,which would allow for more comprehensive training and testing of the capabilities of the two larger models.

\begin{figure}[H]
    \centering
    \includegraphics[width=1.0\linewidth, keepaspectratio]{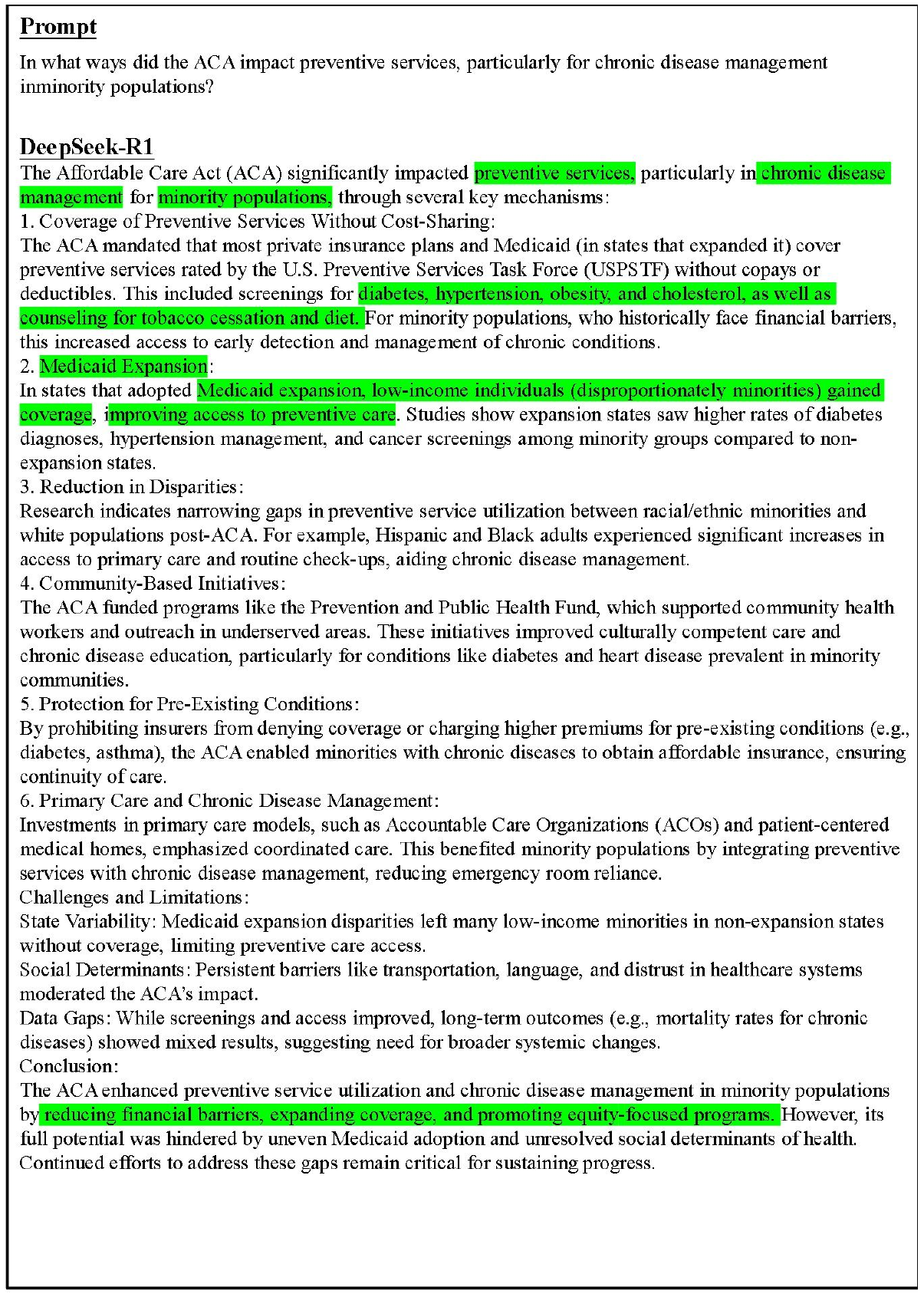}
\end{figure}

\begin{figure}[H]
    \centering
    \includegraphics[width=1.0\linewidth, keepaspectratio]{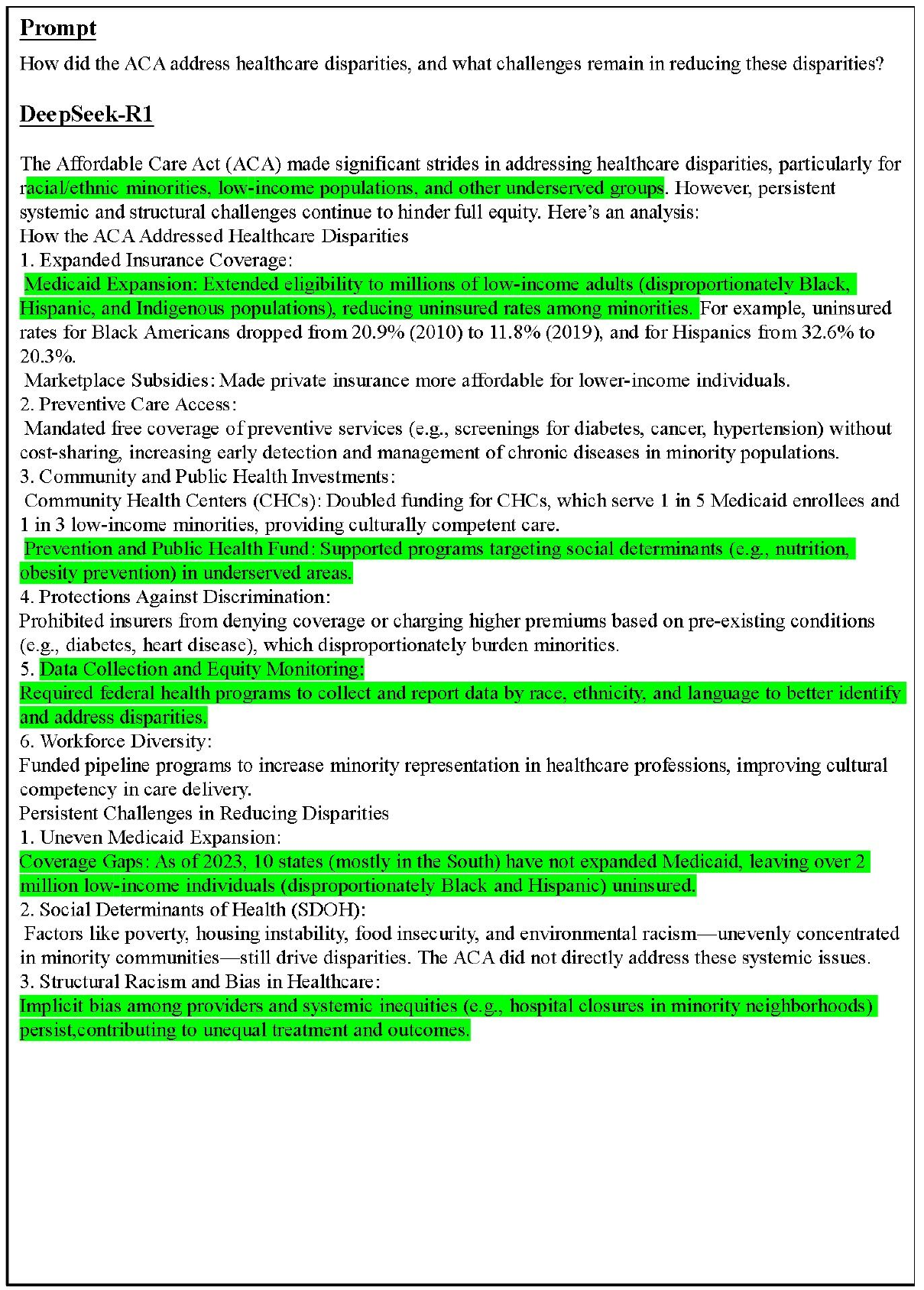}
\end{figure}

\newpage

\subsection{Art Education}

This section analyzes the performance of DeepSeek-R1 in art education and compares the results with those given by o1-preview(The test results of o1-preview come from \textit{Evaluation of OpenAI o1: Opportunities and Challenges of AGI}~\cite{zhong2024evaluation}) and human experts. The test focused on two key issues: explaining the concept of currere in education~\cite{Pinar2019} and  designing a cardboard assemblage art activity for children~\cite{Penfold2020}, analyzing the models' theoretical understanding as well as their analytical and planning capabilities.

\begin{figure}[H]
    \centering
    \includegraphics[width=1.0\textwidth]{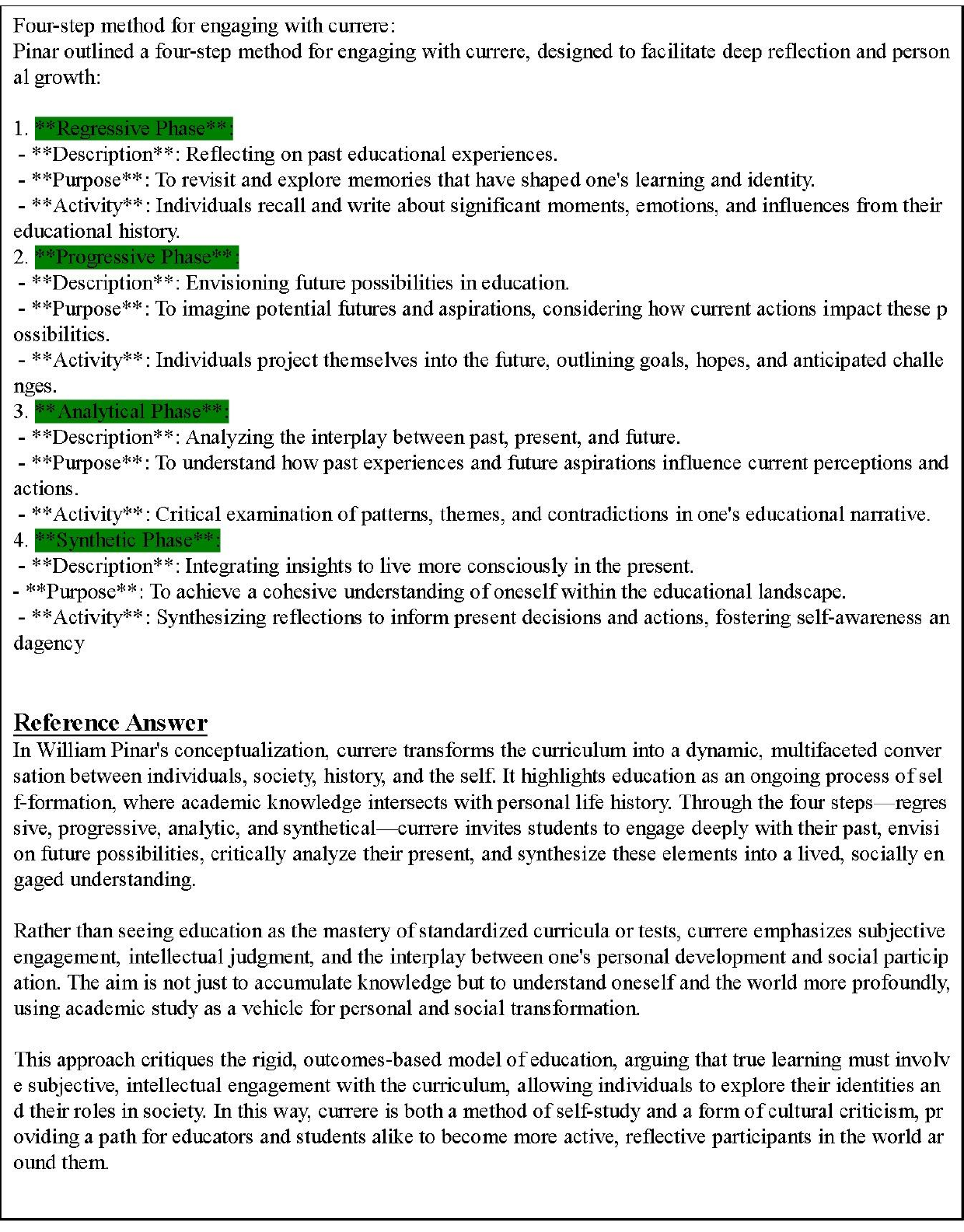} 
    \caption{{\textbf{Art Education: Case 1.} An example illustrating the capability of o1-preview and DeepSeek-R1 to explain the concept of \textit{currere}.}}
    \label{fig:Art_Education1}
\end{figure}

When explaining the concept of currere in education, both DeepSeek-R1 and O1-preview correctly identified the four stages of the curriculum—regressive, progressive, analytical, and synthetic—and analyzed the significance and roles of each stage, emphasizing the importance of personal reflection in education, which aligns with William Pinar's definition of currere~\cite{Pinar2019} .These responses demonstrate DeepSeek and O1 models' accurate grasp of educational curriculum concepts, showcasing their robust capabilities in education-related domains. However, compared to O1-preview's answers, DeepSeek-R1's responses extended beyond mere conceptual explanations. Under the same prompts, it highlighted the origin and source of the concept, discussed its practical implications, and addressed its limitations. This reflects DeepSeek's more comprehensive understanding and critical thinking when analyzing curriculum concepts, as illustrated in (Figure~\ref{fig:Art_Education1}).

\begin{figure}[H]
    \centering
    \includegraphics[width=1.0\textwidth]{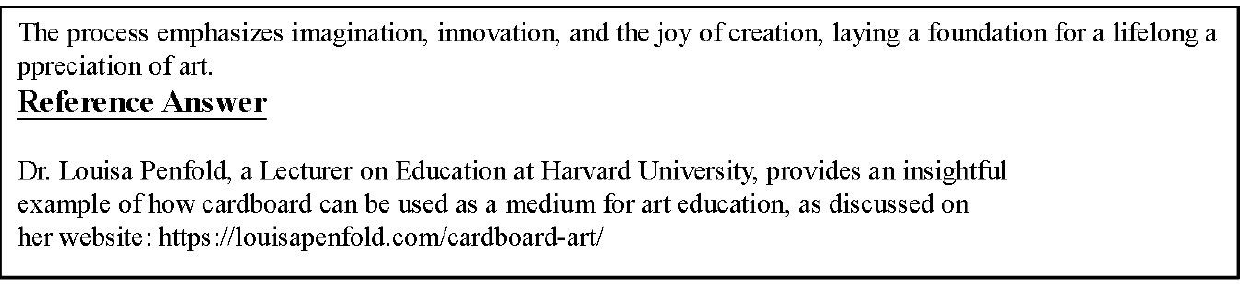} 
    \caption{{\textbf{Art Education: Case 2.} An example illustrating the capability of o1-preview and DeepSeek-R1 to design creative in-class activities for children.}}
    \label{fig:Art_Education2}
\end{figure}

When tasked with designing a cardboard assemblage art activity for children, both DeepSeek-R1 and o1-preview formulated creative objectives to stimulate children’s imagination and provided structured procedural plans, including detailed explanations of material selection and step-by-step instructions. However, compared to the approach proposed by Dr.Penfold—Harvard University’s Art Education Coordinator—their plans appeared overly rigid and formulaic. Dr.Penfold emphasizes allowing children to freely explore, introducing techniques only when necessary, and encouraging diverse problem-solving methods~\cite{Penfold2020}. In contrast, DeepSeek-R1 and o1-preview prioritized structured workflows: first demonstrating examples and techniques, followed by children’s free creation, and concluding with a summary of the project. Compared to human educators, these models still lack flexibility and personalization. Notably, however, DeepSeek-R1’s inclusion of Adaptations, Tips for Success, and Extension Ideas revealed a deeper understanding of the task. By addressing differentiated focuses for various age groups, tolerance for chaotic elements during the activity, and expansions of content, DeepSeek-R1 demonstrated more human-centered considerations in its problem-solving approach, as illustrated 
in (Figure~\ref{fig:Art_Education2}).

In conclusion, both the DeepSeek-R1 and o1-preview models demonstrate strong performance in theoretical interpretation and curriculum planning. However, they still lack the flexibility and sensitivity of human educators when addressing practical curriculum design. Nevertheless, it is undeniable that these models serve as valuable tools for educators, aiding in generating preliminary ideas and establishing foundational frameworks. Compared to o1-preview, DeepSeek-R1 exhibits superior performance in tackling similar challenges, whether through its comprehensive grasp of theoretical concepts or its nuanced considerations during activity planning. This highlights DeepSeek-R1's robust capabilities in addressing complex issues within the field of art education.

\begin{figure}[H]
\centering
\includegraphics[width=1.0\linewidth, keepaspectratio]{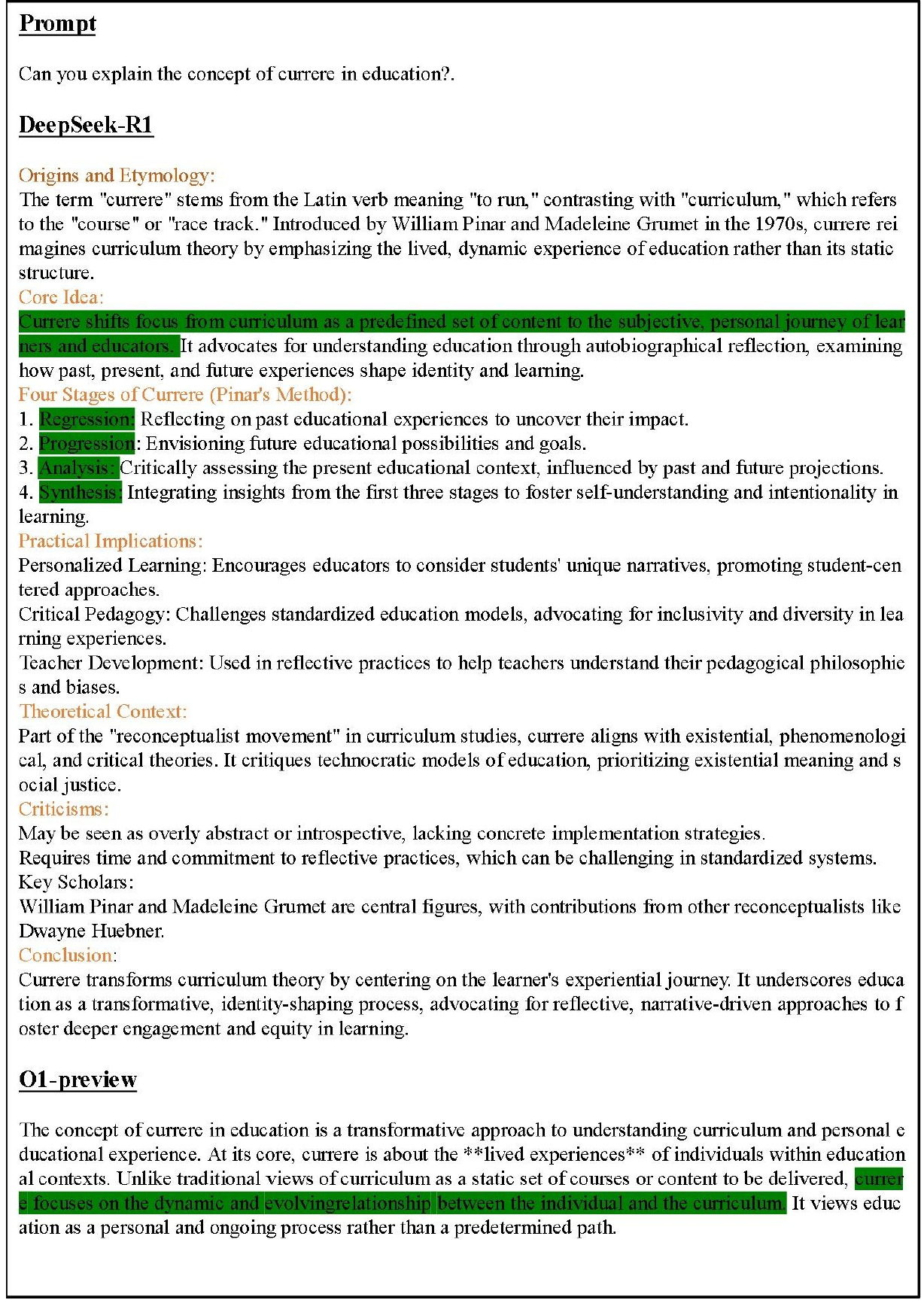}
\end{figure}

\begin{figure}[H]
    \centering
    \includegraphics[width=1.0\linewidth, keepaspectratio]{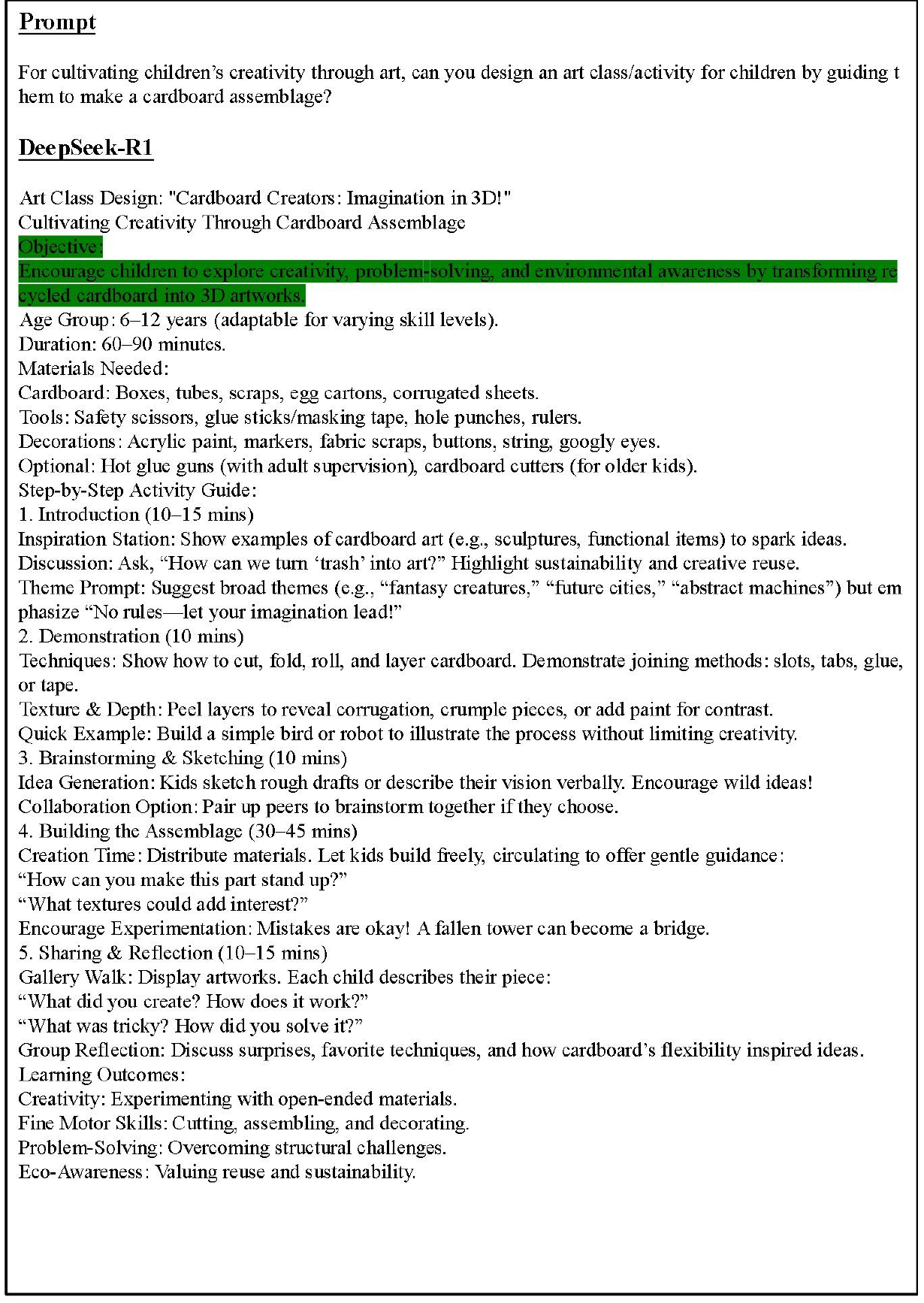}
\end{figure}
\begin{figure}[H]
    \centering
    \includegraphics[width=1.0\linewidth, keepaspectratio]{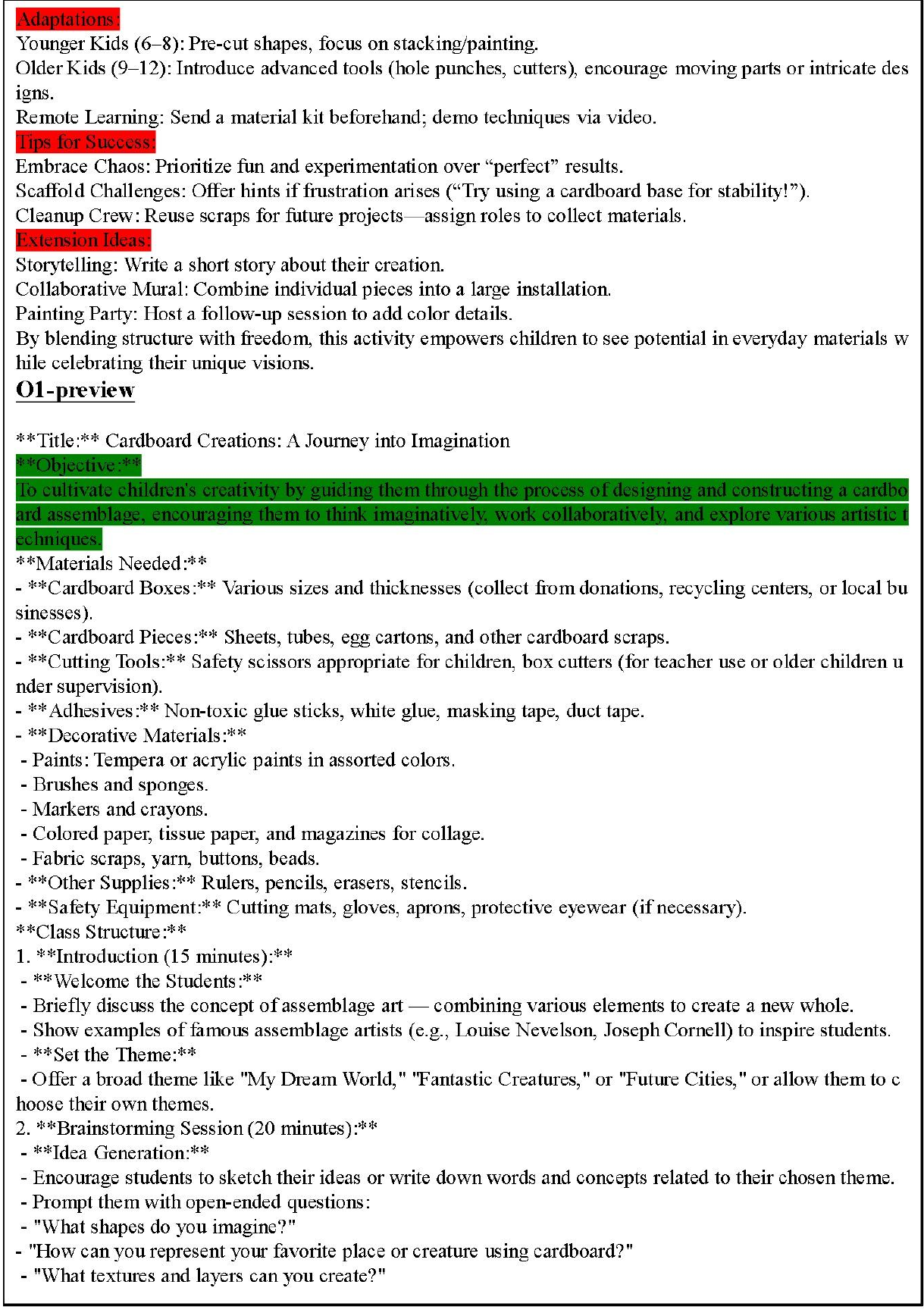}
\end{figure}
\begin{figure}[H]
    \centering
    \includegraphics[width=1.0\linewidth, keepaspectratio]{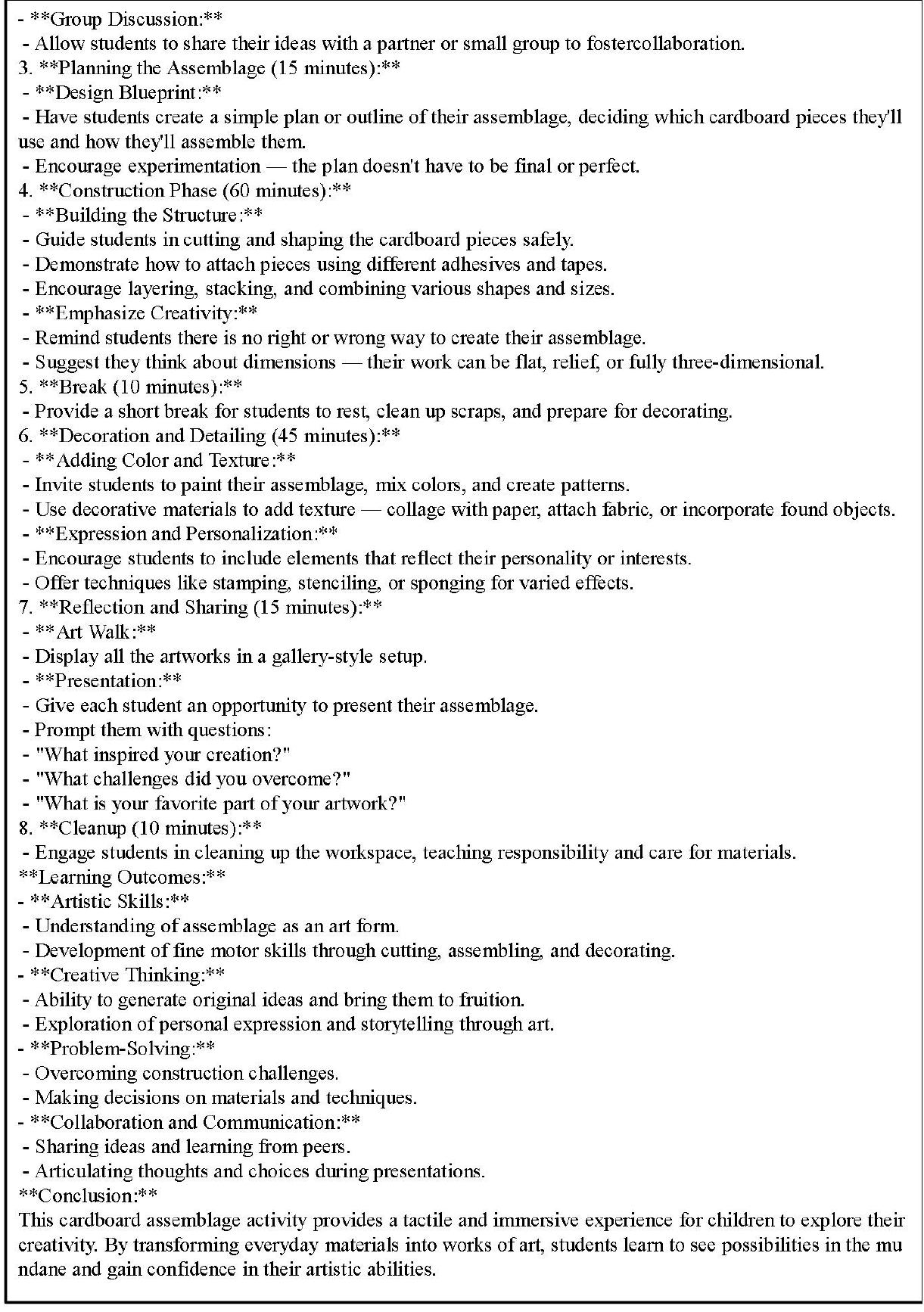}
\end{figure}

\newpage

\section{Conclusion}

This article mainly compares the performance of two newer large language models, DeepSeek-R1 and o1-preview, in seven humanities and social sciences fields: low-resource language translation, educational question-answering, student writing improvement in higher education, logical reasoning, educational measurement and psychometrics, public health policy analysis, and art education. Large language models have great potential in the humanities and social sciences. For example, in social research, LLM can be used for legal and policy text analysis, etc. In the education field, it can be used for AI-assisted student writing improvement, intelligent teaching assistants and personalized learning, etc. In the cultural field, large models can also be used for machine translation and to assist in artistic creation.

In the seven tasks in the humanities and social sciences selected in this paper, the following conclusions were obtained:

\hspace{2em} \textbullet In terms of low-resource language translation, DeepSeek-R1 can perform basic word and grammar recognition for low-resource languages, with similar results to o1-preview, but with slightly different translation styles. Both have their strengths, but both face challenges in dealing with the subtle contextual details and accuracy required for more complex scenarios.

\hspace{2em} \textbullet In terms of student writing improvement in higher education, the content generated by DeepSeek-R1 shows significant advantages in terms of discourse depth, terminology precision, and emotional expression. In contrast, the output of o1-preview tends to be more mechanical and theoretical.

\hspace{2em} \textbullet In terms of educational Q\&A, DeepSeek-R1 demonstrates the powerful capabilities of large language models in the field of educational question answering, which can avoid misleading interference factors and select the correct answer. It is more detailed than the analysis process given by o1-preview.

\hspace{2em} \textbullet In terms of logical reasoning, DeepSeek-R1 can still maintain high reasoning ability and practicality when training samples are limited. Its analysis process is more specific than o1-preview, but not as accurate as o1-preview. The reasoning process of DeepSeek-R1 is innovative, and its conclusions may deviate from traditional answers in some complex situations.

\hspace{2em} \textbullet In terms of educational measurement and psychometrics, DeepSeek-R1 excels in interpretation, logical reasoning, and contextual understanding, while o1-Preview provides accurate responses and excels in concise and efficient calculations.

\hspace{2em} \textbullet In terms of public health policy analysis, DeepSeek-R1 excels in providing detailed analysis, data support, and structured discussion. Although o1-Preview also provides valuable information, it is slightly lacking in detail and comprehensiveness.

\hspace{2em} \textbullet In terms of art education, DeepSeek-R1 performs well in both theoretical interpretation and course planning, but still lacks the flexibility and sensitivity of human educators in actual course design. Compared with o1-preview, DeepSeek-R1 performs well in comprehensive grasp of theoretical concepts and careful consideration during activity planning.

DeepSeek-R1 and o1-preview have a high accuracy rate in the humanities and social sciences, and they also provide excellent answers to open-ended questions. Most of the answers they generate are logically clear and natural in language, and sometimes they can generate creative and logical answers. The two models differ mainly in the form of analyzing questions. For example, DeepSeek-R1 usually gives a point-by-point and longer answer, which is suitable for people who don't know this knowledge quickly to learn, while o1-preview is more inclined to give a one- or multi-paragraph answer suitable for quick reading, which is suitable for people with a certain amount of basic knowledge to verify their own views.

In general, DeepSeek-R1 and o1-preview, as newer large language models, have performed well in the humanities and social sciences. Their analysis processes have their own characteristics and are suitable for reference and learning by different groups of people. DeepSeek-R1 has attracted relatively more attention because of its lower cost while maintaining better results.

\newpage

\bibliographystyle{unsrt}
\bibliography{ref}

\end{document}